\documentclass{article}  

\usepackage{graphicx}

\usepackage{geometry}
\usepackage{txfonts}
\usepackage{amssymb}
\usepackage{amstext}
\usepackage{natbib}
\usepackage[T1]{fontenc}
\usepackage{hyperref}
\usepackage[title]{appendix}
\usepackage{underscore}
\usepackage{tabularray}
\usepackage{array,multirow,booktabs}

\usepackage{caption}
\DefTblrTemplate{conthead-text}{default}{(Continued)}

\DefTblrTemplate{capcont}{default}{%
  \tablename\space\thetable\space
  \UseTblrTemplate{conthead-text}{default}%
}

\SetTblrInner[longtblr]{rowsep=0pt}

\SetTblrTemplate{caption}{empty}

\newenvironment{MYITEMIZE}{\begin{list}{$\bullet$}{
  \setlength{\leftmargin}{15mm}
  \setlength{\rightmargin}{0pt}
  \setlength{\itemindent}{0pt}
  \setlength{\itemsep}{1pt}
  \setlength{\parskip}{0pt}
  \setlength{\parsep}{0pt}}}{\end{list}}

\begin{document} 


\Large
\begin{center}\textbf{Juice-SWI during the Lunar-Earth-Gravity-Assist (LEGA) -- \\
Part 2: Instrument operations}\end{center}\normalsize

\large\noindent T. Cavali\'e$^{1}$, R. Moreno$^{2}$, L. Rezac$^{3}$, F. Herpin$^{1}$, C. Jarchow$^{3}$, P. Hartogh$^{3}$, A. Carrasco-Gallardo$^{3}$, S. Goodyear$^{3}$, P. Mancini$^{1}$, A. Schulz-Ravanbakhsh$^{3}$, B. Dabrowski$^{3}$, Y. Kasai$^{4}$, E. Lellouch$^{2}$, A. Murk$^{5}$, D. Murtagh$^{6}$, M. Olberg$^{6}$, M. Rengel$^{3}$, H. Sagawa$^{7}$, S. Szutowicz$^{8}$, E. Wirstr\"om$^{6}$\normalsize\\
\vspace{0.2cm}

\noindent$^1$Laboratoire d'Astrophysique de Bordeaux, Univ. Bordeaux, CNRS, B18N, all\'ee Geoffroy Saint-Hilaire, 33615 Pessac, France (ORCID: 0000-0002-0649-1192, email: thibault.cavalie@u-bordeaux.fr)\\ 
$^2$LIRA -- Laboratoire d'Instrumentation et de Recherche en Astrophysique, Observatoire de Paris, Section de Meudon, 5, place Jules Janssen -- 92195 MEUDON Cedex\\
$^3$Max-Planck-Institut f\"ur Sonnensystemforschung, G\"ottingen, Germany\\
$^4$Institute of Science Tokyo, Tokyo, Japan\\
$^5$Institute of Applied Physics, University of Bern, Sidlerstrasse 5, Bern, 3012, Bern, Switzerland\\
$^6$Department of Physics and Astronomy, Chalmers University of Technology, 412 96 Gothenburg, Sweden\\
$^7$Faculty of Science, Kyoto Sangyo University, Kamigamo-Motoyama, Kita-ku, Kyoto 603-8555, Japan\\
$^8$Centrum Bad\'an Kosmicznych Polskiej Akademii Nauk, Bartycka 18A, Warsaw, Poland\\

\vspace{0.2cm}
\noindent\textbf{Received:} 31 January 2026\\
\noindent\textbf{Accepted:} 9 April 2026\\
\vspace{0.2cm}

\noindent\textbf{DOI:} 10.5194/angeo-44-461-2026 \\
\vspace{0.5cm}

\section*{Abstract}
The Jupiter Icy Moons Explorer (Juice) embarked in 2023 on a 8-year interplanetary journey to Jupiter and its icy moons. The Submillimetre Wave Instrument (SWI) is one of the ten science instruments aboard the spacecraft. SWI is a sophisticated and first-of-its-kind payload visiting the outer solar system, featuring dual-band tunable receivers, two independent pointing mechanisms, and spectrometers capable of high resolution (up to a resolving power of 10$^7$). It is designed to support the diverse science objectives of the Juice mission targeting Jupiter's middle atmosphere, icy-moon's exospheres as well as near sub-surface thermophysical properties. For this purpose the Juice mission adopts a complex trajectory tour within the Jovian system, which further necessitates a sophisticated, mission-driven operations concept for SWI. This presents significant planning, operations and commanding challenges which are described in this paper in the context of the Lunar and Earth Gravity Assist (LEGA). After the development and ground calibration of the instrument, the SWI Team has designed a comprehensive calibration strategy applicable during the Cruise Phase of Juice. Among the various opportunities for calibration, including the Near-Earth Commissioning Phase and more than ten Payload Checkout Windows, the LEGA offers the means not only to improve the calibration of the instrument, but also to validate the operational strategy of future icy moon flybys. 

\section{Introduction}
The Jupiter Icy Moons Explorer (Juice, \citealt{Grasset2013}) is a mission chosen in the framework of the Cosmic Vision 2015-2025 programme of the Science and Robotic Exploration Directorate of the European Space Agency (ESA). Juice was launched on April 14, 2023, from the Kourou space port (French Guiana) and has since embarked on a 8-year interplanetary transfer to Jupiter, the so-called Cruise Phase. The Cruise Phase is the main mission phase for commissioning, in-flight performance monitoring and calibrating the instruments. Within the three months following launch, the Near-Earth Commissioning Phase (NECP) took place. Payload Checkout Windows (PCWs) are then scheduled on a regular basis and allow instrument teams to check their instrument health for one week every semester on average. The various Earth swingbys do not only offer additional time windows to complement the instrument verification, but they are also unique opportunities to test multi-day planning and observations. The first planetary swingby of the Juice Cruise Phase was the Lunar-Earth Gravity Assist (LEGA), which occurred in August 2024. It will be followed by two other Earth Gravity Assists (EGA) planned in September 2026 and January 2029. In July 2031, Juice will perform its Jupiter Orbit Insertion (JOI) that will start its Jupiter orbital tour. According to the current baseline trajectory \citep{Boutonnet2024}, Juice will then perform a series of Jovian orbits (equatorial and inclined) with 62 perijoves along with 36 moon flybys (2 of Europa, 11 of Ganymede, 23 of Callisto), before performing an orbital insertion around Ganymede (GOI) in December 2034. The final leg of the Juice mission will consist of a suite of elliptical and circular orbits around this icy moon until the end of the nominal mission in late 2035. The Ganymede circular orbits (GCO) will include high altitude (5000 km), low altitude (500 km), and very low altitude (200 km) orbits.  

The Submillimetre Wave Instrument (SWI, \citealt{Hartogh2026a,Hartogh2026b}) is one of the ten science instruments aboard Juice. SWI will study the composition, chemistry and dynamics of the atmospheres of the Galilean satellites, as well as their surface properties. It will also investigate the chemistry, dynamics and structure of Jupiter's middle atmosphere and the coupling processes between the planet's atmosphere, its magnetosphere and interplanetary environment. By means of high-resolution (up to R$=$10$^7$) line spectroscopy and continuum radiometry, it will:
\begin{MYITEMIZE}
  \item Characterize, in a unique and unprecedented manner, the tenuous atmospheres / exospheres of the Galilean satellites \citep{depater2023a,Wirstrom2020}, enabling determination of their sources and sinks (e.g., \citealt{Marconi2007}), and interactions with the Jovian magnetosphere,
  \item Measure surface / subsurface properties of icy satellites, constraining thermo-physical properties (e.g., thermal inertia, dielectric constant, porosity) and the surface-bounded atmosphere spatial distribution, composition, and dynamics in relation to the exospheres of these bodies \citep{deKleer2021},
  \item Provide a detailed characterization of the thermal field, dynamics, and composition of Jupiter's stratosphere to constrain its general circulation \citep{Cavalie2021,Cavalie2023b}, and the coupling of the stratosphere with the lower and upper atmosphere \citep{Medvedev2013,Guerlet2020,Boissinot2024},
  \item Determine key isotopic ratios in Jupiter's and the satellites' atmospheres, constraining the origin and evolution of the jovian system \citep{Mousis2014b,Gapp2024,Lefour2026}.
\end{MYITEMIZE}

SWI is a heterodyne radiometer/spectrometer operating in two different submillimetre channels using two independent local oscillators (LO) to observe the sky simultaneously in two wavelength bands, 233-281\,$\mu$m (1066-1286\,GHz) and 470-565\,$\mu$m (530-638\,GHz), with selected spectrometers. It can record the sky emission over two spectral windows, either with very high spectral resolution Chirp Transform Spectrometers (CTS, 1\,GHz bandwidth, 10000 channels, 100\,kHz resolution), the high resolution Auto-Correlation Spectrometers (ACS, 4.4\,GHz bandwidth, 1024 channels, 4.3\,MHz resolution), or with the Continuum Channels (CCH, 4\,GHz bandwidth). The CCH can be used as stand-alone or in combination with either the ACS or the CTS. SWI is a miniaturized observatory aboard Juice, comprising a Telescope and Receiver Unit (TRU), an Electronic Unit (EU), and a radiator. The TRU has a main antenna with a diameter of 29\,cm. The TRU is equipped with two orthogonal pointing mechanisms that allow to point to any position in the sky from the nadir direction of the Juice platform with a freedom of $\pm$72$^\circ$ along track (AT) and $\pm$4.3$^\circ$ cross track (CT) during the Jupiter orbital phase of the Juice mission. It also bears a flip mirror (FLM) that enables switching from an optical path pointing to the sky to another path pointing to an internal calibration hot load. 

SWI is thus a complex instrument that requires a comprehensive calibration plan, similarly to other heterodyne instruments flown on submillimetre observatories, like the Heterodyne Instrument for the Far-Infrared of Herschel \citep{Roelfsema2012} or the receivers of Odin \citep{Olberg2003}. This plan aims to understand the instrument's stability, spectral and spatial responses, enable proper radiometric calibration of the data, etc., and to track the temporal evolution of all these parameters. It is described in \citet{Hartogh2026a}. SWI stands out uniquely in its complexity from the science plan perspective. Unlike missions/instruments such as Odin, Herschel, or Rosetta/MIRO (Microwave Instrument for the Rosetta Orbiter), where target and orbital scenarios remained relatively stable, the science phase of SWI involves dynamically changing targets and orbits, and science goals. This evolving operational context introduces a level of challenge and adaptability that was not present in previous missions, making the SWI planning and execution significantly more intricate. To execute the calibration and science plan, the sub-units of the instrument need to be commanded by its Digital Processing Unit (DPU) in order to build calibration observations and observations of science targets. Section \ref{sec:observation_modes} presents the concept and description of the observation modes which are used to operate the instrument. Section \ref{sec:conops} details the concept of SWI operations planning. Finally, section \ref{sec:obs_strategy} exposes the overall observation strategy of SWI for the Cruise Phase.

\section{Observation modes \label{sec:observation_modes}}
SWI operates with two hierarchical levels of modes. First, there are the instrument modes, a term which is used to communicate the highest level of controlling of the instrument with the onboard application software. Second, there are the observation modes. The observation modes, as will be discussed, are encapsulated and can be also referred to as ``scripts''/``science scripts''. They can be invoked with parameters when the instrument is running in its ``science'' instrument mode. These two types of modes are described in this section.

  \subsection{Instrument modes}
  There are 7 basic instrument operational modes that can be commanded with the SWI application software:
  \begin{MYITEMIZE}
    \item OFF: All instrument subsystems including the DPU are switched off. Consequently, there is no housekeeping data and no telemetry generated by the instrument itself. The instrument is in this mode during launch and Cruise Phase, except during calibration campaigns (e.g., NECP, PCWs, Earth and Moon flybys).
    \item STANDBY: Only the instrument DPU is switched on and is able to receive instrument commands. Only housekeeping telemetry is generated in this mode.
    \item SAFE: The DPU and ultra-stable oscillator (USO) are switched on. This mode is used in different cases. It serves for the USO stabilization prior to warm-up. It is also the mode into which the instrument switches automatically in case an instrument anomaly is detected. SWI is also switched to this mode at the end of science operations and is meant to be used during downlink. Only housekeeping telemetry is generated in this mode.
    \item DIAGNOSTIC: The DPU, USO, and selected spectrometers (CTS or CTS/CCH or ACS or ACS/CCH) are switched on. Diagnostic activity is allowed in this mode, including activation and control of sub-units.
    \item UNLOCK: Only the DPU is switched on. This mode was only used once post-launch to release the launch locks of the instrument mechanisms that kept the TRU in its launch position.
    \item WARMUP: The DPU, USO, and selected spectrometers (CTS or CTS/CCH or ACS or ACS/CCH) are switched on. This mode is a transition mode to enter the SCIENCE mode.
    \item SCIENCE: The DPU, USO, and selected spectrometers (CTS or CTS/CCH or ACS or ACS/CCH) are switched on. This mode serves to invoke calibration and science observations with the appropriate list of mode parameters (see next Section).
  \end{MYITEMIZE}

\begin{figure}[t]
    \centering
    \includegraphics[width=14cm]{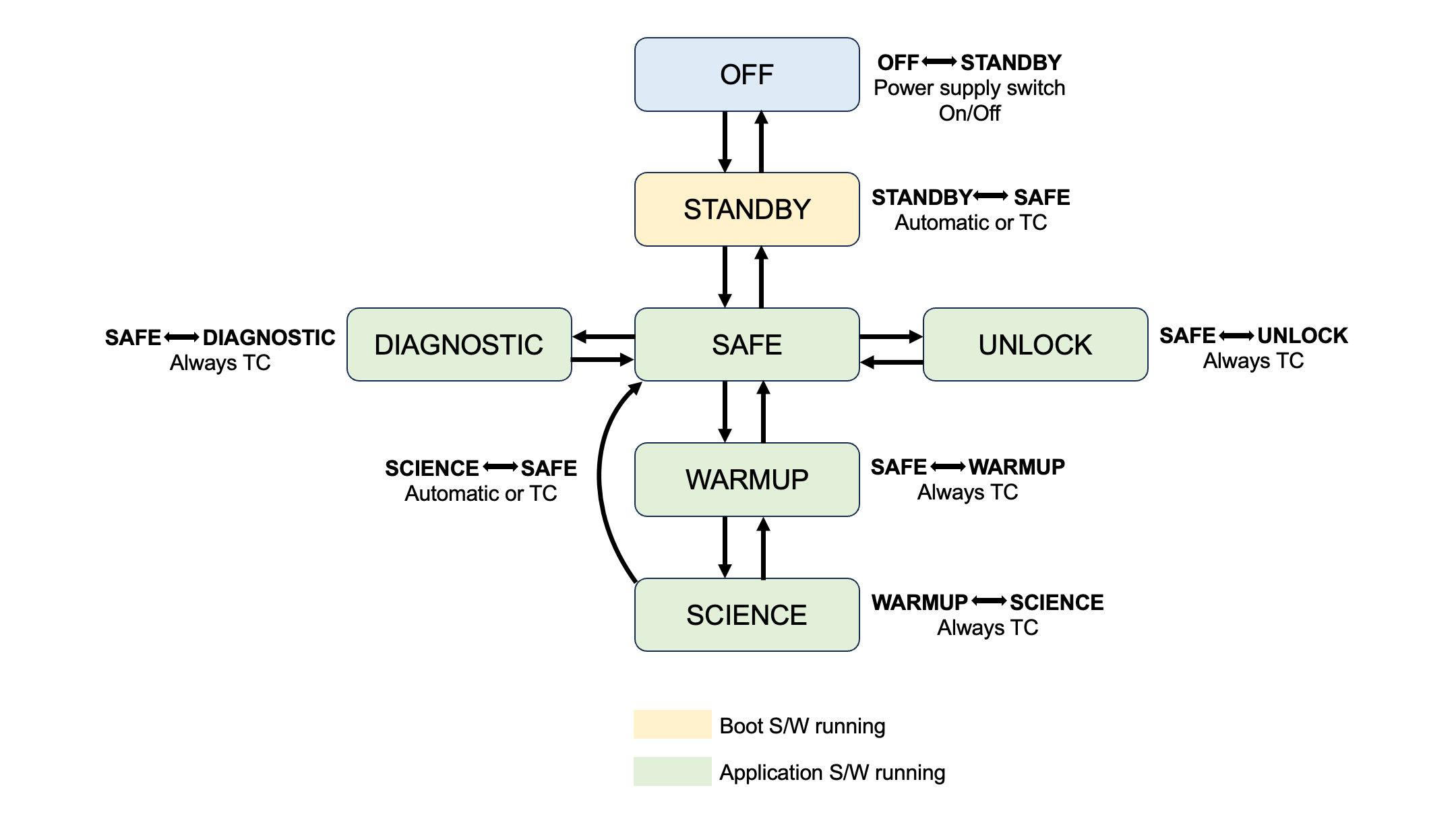}   
    \caption{Transitions between instrument modes in the SWI DPU. Science observations are performed when the instrument is set to its SCIENCE mode. Credit/property of MPS. }
    \label{fig:transitions}
  \end{figure}

  The allowed transitions between these instrument modes are illustrated Fig.~\ref{fig:transitions}.

  \subsection{Concept of observation modes}
  During the Science Phase (Jupiter tour and Ganymede orbits), Juice will operate on the basis of 16 hrs of science operations per day, followed by 8 hrs for communications with the Earth ground stations (downlink and uplink activities). Because of telemetry budget limitations, ESA has defined the number of 100 telecommands (TC) as the limit for uplink per science day per instrument. This limit makes it impossible to operate SWI with individual sub-unit commands, because a single science observation lasting just a few minutes would already require tens to hundreds of those commands. An observation indeed requires several steps to be taken, like the tuning of the receivers, setting the pointing mechanism, pointing to the science target, performing the required ON-OFF integrations (including pointing to the cold sky for OFF integrations), and making the relevant hot-cold calibration integrations (including the relevant FLM movements). Even the simplest observation, i.e., a nadir stare without any repetition of the integration loop, requires a minimum of 16 commands. Early in the development phase of the instrument, it was therefore decided by the instrument team to adopt the concept of observation modes, similarly to Herschel-HIFI \citep{Roelfsema2012}. These are a set of scripts, pre-loaded in the DPU memory, that can be invoked by single TC, once the instrument is set in the SCIENCE Mode. Each script is a suite of sub-unit commands, which can be grouped in repetition loops, in order to produce a complete observation (e.g., 2D map). The details of the operation of each single script execution are then uniquely governed by the list of parameters supplied by the TC. 

  \subsection{General description \label{sec:obs_modes}}
  SWI currently has a total of 34 observation modes implemented in its DPU. Observation modes can be grouped in two categories: calibration modes meant for characterization and monitoring of the instrument and its behavior, and science observation modes meant for science measurements. A mode/script is fully defined by a set of parameters which concerns, for example, tuning, receiver and spectrometer setup, mechanism movements for pointing. All modes are provided with details in the Appendices \ref{sec:appendix_calibration} and \ref{sec:appendix_science}.
  
    \subsubsection{Calibration modes}
    There are 9 calibration modes for SWI. They have been designed to characterize the Allan variance and to measure the system temperature with all backends (CTS, ACS, CCH). The Allan variance modes (all modes with names starting with ``SWI\_ALLAN'', see Appendix \ref{sec:appendix_calibration}) use the shortest integration time possible and accumulate integration time in order to characterize up to which integration time the spectral noise decreases according to the radiometric formula \citep{Schieder2001}. It can be applied to any of the possible tunings of the instrument. The radiometric formula links the spectral noise with the integration time of an observation\footnote{It should be noted that there is no direct known effect of radiation on the signal quality (not more noise) of SWI, so it is not expected that the sensitivity of SWI suffers from the Jovian radiation environment.} and requires the knowledge of the system temperature \citep{Tolls2004,Gulkis2007}. The system temperature modes (all modes with names starting with ``SWI\_TSYS'', see Appendix \ref{sec:appendix_calibration}) consist of a spectral scan across several tunings that can be supplied via the TC. For instance, the system temperature mode using the CTS can cover up to 15 tunings in each band. The SWI mechanisms may be used to point SWI beam to the cold sky at the start of all these calibration modes. However, the mechanisms are not used during the actual measurements.

    \subsubsection{Science observation modes}
    To cover all the science objectives of the instrument, the SWI Team has defined a total of 22 science observation modes. The science observation modes are distinct by their pointing pattern, which have been carefully designed to match all the observation requirements derived from the instrument science goals \citep{Hartogh2026a}. All modes contain repetition loops, i.e., sequences that enable the accumulation of integration time on the selected target (single point or maps). 

    The generic science modes enable the observation of a target either with: (i) a nadir stare (all modes with names starting with ``SWI\_NADIR\_STARE'', see Appendix \ref{sec:appendix_science}), with the possibility to define an offset to the nadir position to aim for a particular latitude/longitude/region-of-interest of the target (e.g., \citealt{Stephan2021} for Ganymede), (ii) a 5-point cross for low spatial resolution mapping (all modes with names starting with ``SWI\_5POINT\_CROSS'', see Appendix \ref{sec:appendix_science}), (iii) a 2D map consisting of evenly separated rows and columns (all modes with names starting with ``SWI\_2D\_MAP'', see Appendix \ref{sec:appendix_science}). The 2D maps are very versatile and can be customized to many different purposes by setting the proper  parameter values. For instance, the 2D map mode can serve to perform a zonal (resp. meridional) scan by setting the number of rows (resp. columns) to one. Also, the separation between two consecutive rows can differ from the separation between two consecutive columns in a 2D map. The observation principle of these modes is shown in Figs.~\ref{fig:Nadir_stare} and \ref{fig:2D_map}. 

    Most of the science observation modes are defined for use with the CTS simultaneously with the CCH. All these modes enable single or (preferentially) dual band observations of a target, by defining one or two tunings. They can also be used either based on the position-switching calibration technique, where the instrument alternates between the science target and the cold sky to achieve radiometric calibration of the measurements, or based on the frequency-switching technique, where the LO is tuned to a nearby frequency to perform the radiometric calibration. The latter is generally used for the observation of narrow lines \citep{Hartogh2010a,Biver2015} and enables spending nearly 100\% of the integration time on the target, provided that the spectral lines are significantly narrower than the LO frequency throw applied in the scheme. The 2D map mode also comes in an on-the-fly version (all modes with names starting with ``SWI\_2D\_MAP\_OTF''), in which a cold sky measurement is taken only once at the end of each row of the map (\citealt{Ossenkopf2009}, and see examples of applications in \citealt{GallardoCava2023} and \citealt{Szabo2025}). The cold sky measurement ending a row is then applied to all the on-source points of that same row, instead of being recorded after each on-source point of the row when using the position-switching version of the mode. The duty cycle is then largely improved, provided that the instrument is stable enough (i.e., the instrument gain drifts are negligible) over the duration of a row. 

    In addition, spectral scan modes (all modes with names starting with ``SWI\_SPECTRAL\_SCAN'', see Appendix \ref{sec:appendix_science}) have been defined for use either with the CTS or with the ACS, either with the position-switching calibration scheme or with the frequency-switching scheme. These spectral scans enable to observe many tunings in one execution (see example in \citealt{Costagliola2015,DeBeck2018}), for instance, 15 tunings for the spectral scan with the CTS and position-switching. Similarly to the nadir stare modes, the spectral scans are nominally nadir, but can be offset to any region of interest of a target (e.g., limb direction).

    Using the capabilities of SWI, target-specific science observation modes have also been designed to maximize the science return of the Jupiter and moon observations. One of the main science goals of the instrument is to constrain the general circulation in the stratosphere of Jupiter. It is thus crucial to measure temperature and winds preferably simultaneously with the required accuracy to constrain the mechanical forcings at play in the atmosphere \citep{Hartogh2026a}. This is best achieved with limb observations of the CH$_4$ lines at 1256\,GHz and a strong line in the 600\,GHz band, like the H$_2$O line at 557\,GHz. Alternately, a strong H$_2$O line can be used in the 1200\,GHz band. However, the vertical range over which the temperature can be retrieved is then more limited, because H$_2$O is restricted to the upper and middle stratosphere by condensation \citep{Moses2005,Cavalie2008c}. In all cases, the retrieval of the winds necessitates to know where the instrument is pointed to enable the subtraction of the beam-convolved planet rotation with an accuracy better than the wind accuracy required to constrain the models \citep{Cavalie2021,Benmahi2022,Benmahi2025,Carrion-Gonzalez2023}, because the Doppler shift induced by typical stratospheric winds of 100\,m.s$^{-1}$ is superimposed on a much larger Doppler shift induced by the 12.5\,km s$^{-1}$ planet rotation. To achieve such accuracy in the limb pointing determination, each limb integration with the CTS is preceded by a rapid scan across the limb with the CCH. After completing an OFF measurement, the CCH beam will be placed on the Jovian disk near the targeted limb position. This ``nadir'' CCH measurement will be proportional to the temperature of the atmospheric layers producing the continuum. On Jupiter, the continuum observed in the submillimeter is produced at $\sim$500\,mbar \citep{Cavalie2008c}, where the temperature is around 140\,K. This is slightly above the cloud deck probed at e.g. longer millimeter wavelengths \citep{dePater2019}. While there may be some temperature variability over ranges of $\sim$20$^\circ$ in longitude for a given latitude (and it is a goal of SWI to quantify them), the reference ``nadir'' pointing near the limb should have very similar continuum temperature  (within $\sim$1\,K, i.e. <1\% difference)  as the longitude of the limb. As soon as the beam will start reaching the limb and cold space, the CCH signal\footnote{The inflight sensitivity of the CCH is still under investigation to confirm the CCH noise is sufficiently low not to compromise the across limb scan concept of this observation mode.} will start to drop toward the OFF measurement value. When the recorded signal reaches 50\%~of the (nadir+OFF) value, which then corresponds to a beam filling-factor of 50\%, the instrument boresight will be at the limb. The scan is thus stopped and the integration with the CTS starts. This can be programmed either for a single latitude (all modes with names starting with ``SWI\_JUP\_LIMB\_STARE'', see Appendix \ref{sec:appendix_science}) or for a series of latitudes (all modes with names starting with ``SWI\_JUP\_LIMB\_RASTER'', see Appendix \ref{sec:appendix_science}). The observation principle of these modes is shown in Fig.~\ref{fig:Limb_stare}. These modes are also to be used for deep integrations that aim at detecting isotopes and trace species, because an accurate pointing at the limb maximizes line contrast, which is especially crucial for faint lines. For temperature/wind measurements, the position-switching version of the mode will be used, because the targeted lines are broader than the frequency throw enabled in the frequency-switching modes. For isotopic lines and the search for new species, the frequency-switching version may be used, provided that the temperature information is not needed (because already obtained for the pointed latitude/longitude).

    For the investigation of Galilean moons subsurface, atmospheric abundance, composition and temperature, there are specific nadir stare modes (all modes with names starting with ``SWI\_MOON\_NADIR\_STARE'', see Appendix \ref{sec:appendix_science}). The only difference with the generic nadir stare mode described above is the addition of CCH measurements recorded in parallel of the CTS ones. During a given CTS integration (generally 10-60\,s), CCH measurements are taken in parallel, with a much shorter integration time (generally 0.4\,s), and thus with at a higher cadence. This significantly improves the spatial sampling of the surface/sub-surface temperature, and thus facilitates moon surface science from those continuum measurements, without impacting significantly the total data volume. For winds, temperature, isotopes and trace species, there are two types of limb modes. For deep integrations, one can use the limb stare mode (all modes with names starting with ``SWI\_MOON\_LIMB\_STARE'', see Appendix \ref{sec:appendix_science}), that enables to point at a given latitude and altitude. For temperature, composition and wind vertical profile sampling when the beam is smaller than the apparent width of the atmospheric limb, limb scanning will be preferred (all modes with names starting with ``SWI\_MOON\_LIMB\_SCAN'', see Appendix \ref{sec:appendix_science}), similarly to how \citet{Guerlet2018} operated Cassini/CIRS (Composite Infrared Spectrometer) on Saturn. This mode enables to target a latitude and to scan up and down from the surface to the top of the atmosphere, with an altitude step that is adaptable to the science goal. The observation principle of both limb modes is shown in Fig.~\ref{fig:Moon_Limb}.

  \subsection{Validation of the observation modes}
  The details on how to execute a given science observation have been defined partly based on theory and partly on experience from other microwave or infrared instruments in space (e.g., \citealt{Roelfsema2012}). It is necessary first to verify if the scripts are functional on the ground and then to test inflight if these observation modes provide science data with the expected quality. This mode validation thus constitutes part of the Cruise Phase operations of SWI.

  \subsection{Summary on observation modes}
  SWI is an instrument aiming to observe surfaces and atmospheres in the Jovian system to measure composition, temperature and wind speeds. To achieve all these goals, the instrument was designed with all the flexibility (and complexity) as that found in ground-based sub-mm observatories, including, for instance, pointing pattern, spectral tuning, coverage and sampling, integration times, integration repetition factors, etc. All the calibration and science observation modes described in Section \ref{sec:obs_modes} can be fed with the appropriate parameters to match a given measurement goal. Each observation is thus unique and must not only use one of the modes with the appropriate list of parameters, but it must also carry a series of metadata that provide the necessary information to enable meaningful data reduction and analysis. The description of the concept of operations of SWI is given in the next section.

\section{Observation planning during the LEGA \label{sec:conops}}
  \subsection{Generalities on observation planning}
  The planning of operations during any of the Cruise Phase opportunities, including the LEGA, follows several steps. In a first step, the Mission Operation Center (MOC) defines the boundary conditions of the operations: date/time, duration, allowable data volume, etc. Instrument teams are then asked for a preliminary Activity Plan (APL) with operation names and descriptions, estimates of number of TC, duration and data volume, and applicable constraints. MOC usually proposes a preliminary breakdown of the data volume envelopes per instrument. In a second step, instrument teams submit an initial sequence of operations, in the form of a preliminary list of Payload Operation Requests (POR). At this stage, there can be trade-offs between instruments, managed by MOC. Once the final envelopes are agreed upon, the teams are requested to submit their final PORs several weeks before execution of the operations. 

  During the Science Phase, iterations with instrument teams are managed by Science Operation Center (SOC, \citealt{Altobelli2026}). SOC initially provides a segmentation of the trajectory, which prioritizes operations on a science-driven basis. SOC then handles the various instrument team deliveries, starting with the Observation Plan (OPL). The OPL is a list of instrument observation requests, which are backed up by a science plan. Instrument teams can request observations under two statuses: prime or rider. The prime status implies the instrument has constraints on the spacecraft pointing, while the rider status bears no such implication. After harmonization of the OPLs from all the instrument teams, a skeleton spacecraft Pointing Timeline Request (PTR) is generated and SOC designates which instrument team has to design the spacecraft pointing for each time block of the timeline. Once the final PTR is produced, SOC provides a skeleton Instrument Timeline (ITL), which provides the information on spacecraft resources (power, data rate and data volume). Some operations can then still be removed to comply with, for example, the total power available to instruments at a given point on the timeline.

  Because the LEGA operations were managed by MOC, and because there was no spacecraft pointing allowed during the whole LEGA timeline, the setting of the whole timeline of operations did not necessitate to go through the various steps that will be required in the Science Phase. Some trade-offs had to be found for the closest approaches to the Moon and the Earth, but, as will be discussed in the next sections, SWI was one of the few remote sensing instrument capable of performing observations of the Earth and the Moon in the days that followed the closest approach to the Earth, by using its pointing mechanisms.

  \subsection{Specificity of SWI observation planning}
  The planning of SWI observations requires accurate pointing information that relies on the interpretation of the mission SPICE (Spacecraft, Planet, Instrument, C-matrix, Events) kernels (\href{https://www.cosmos.esa.int/web/spice/spice-for-Juice}{https://www.cosmos.esa.int/web/spice/spice-for-Juice}, last access: 23 April 2026), which are regularly updated and delivered by ESA. This is true for all modes except the calibration modes.

  Geometrical calculations derived from the SPICE kernels must be performed individually for each script in order to obtain accurate pointing information for the observations to be scheduled. Other target-specific geometric information that cannot be derived a posteriori from the commanding parameters such as, for instance, target name, target size, sub-spacecraft latitude and longitude, and local time, must also be included in the metadata of an observation.

\begin{figure}[!t]
    \begin{center}
      \includegraphics[width=15cm]{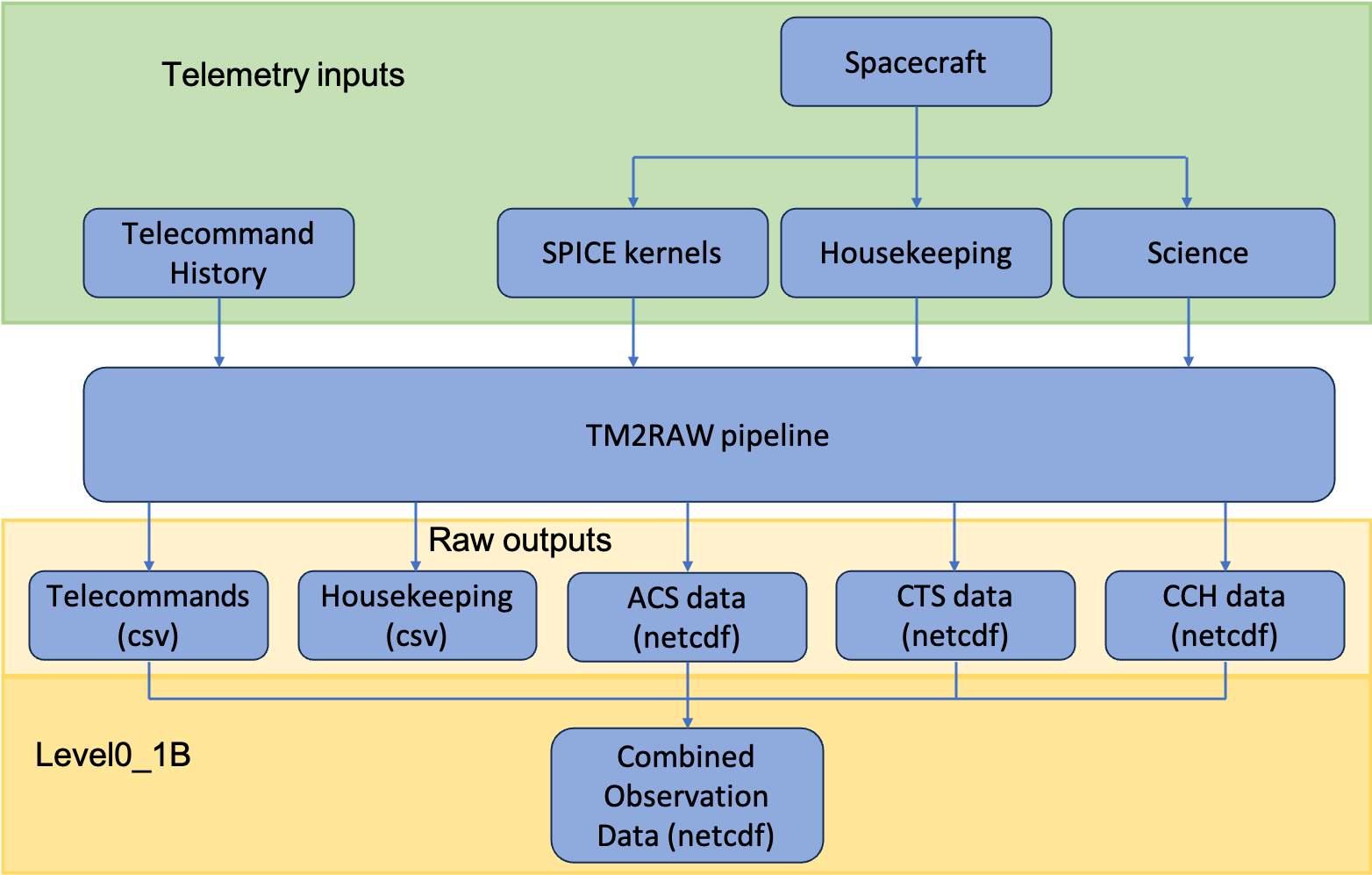}
    \end{center}
    \caption{Flowchart representing the SWI processing of spacecraft telemetry packets into raw data files, and subsequently into Level0\_1B which additionally include the necessary telecommanding metadata. This is a high-level outline of the SWI telemetry-to-raw (TM2RAW) pipeline. The formats of the various output data produced by the TM2RAW pipeline are specified within parentheses. The Level0\_1B files are then inputs into the SWI calibration pipeline.}
    \label{fig:Pipeline} 
  \end{figure}

  In practice, the metadata of an observation are not uplinked to the spacecraft. Only the list of TCs containing the sequence of observation modes with their relevant list of parameters is sent for operation execution. For a given observation, those metadata, the selected observation mode and the parameter list are kept in a database of planned observations. They are used upon the downlink of the telemetry binary files (both housekeeping and science data) by the SWI telemetry-to-raw (TM2RAW) pipeline to automatically create the proper file structure and to populate it with the raw data. A high-level description of the SWI TM2RAW pipeline is shown in Fig.~\ref{fig:Pipeline}. This is absolutely necessary, because, for example, the binary data file from a 60-minute 2D map with 10$\times$5 on-source points with 5\,s integration time per point will have a different structure and contain a different amount of data (e.g., number of spectra) compared to a 10-minute nadir stare. A second implication is that each calibration and science observation mode must have its dedicated data reduction pipeline. It is planned for the near future that the SWI telescope pointing information, although transcribed in the science metadata will also be retrievable from dedicated instrument attitude SPICE kernels.

  Consequently, the SWI Science Team has opted for a unique observation identification number for each planned observation, similarly to what was made for Herschel observations (e.g., \citealt{Hartogh2010b}, \citealt{Cavalie2013,Cavalie2014}). These so-called ObsIDs (observation identification numbers) serve this purpose of tailoring an observation, defined by its observation mode and list of parameters, with its mandatory metadata from the planning stage down to the data reduction stage. They uniquely identify, characterize and naturally encapsulate the SWI data/observations. We should also highlight here that SWI data will be calibrated and archived in chunks identified by the ObsIDs.

  \subsection{The SWI Observation Planning Tools}
\begin{figure}[!t]
    \begin{center}
      \includegraphics[width=15cm]{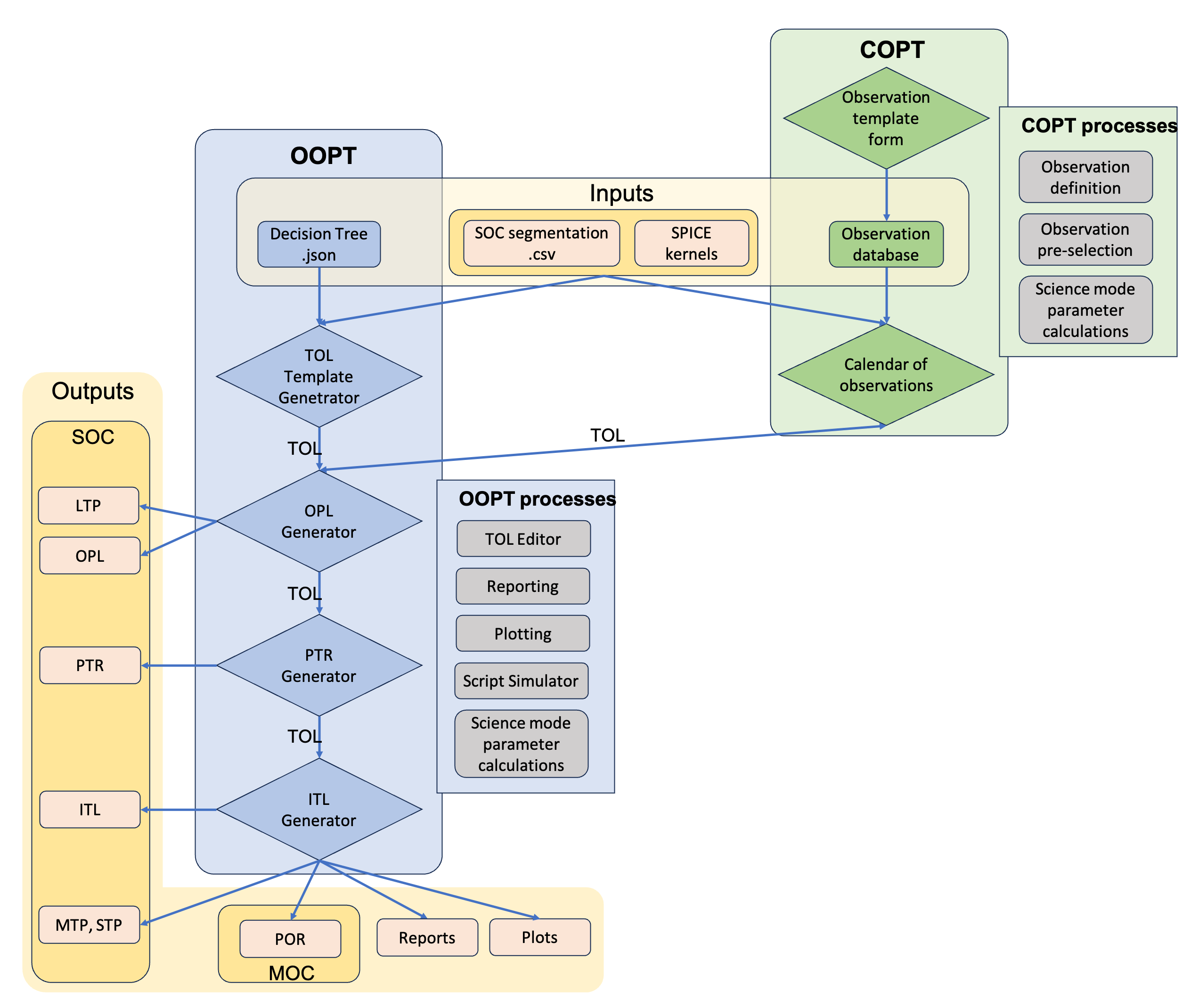}
    \end{center}
    \caption{Flowchart representing the various steps required from initial inputs (e.g., SOC segmentation, SPICE kernels) to the various deliveries (e.g., MTP, LTP, PORs) in the Science Phase. The Science Co-Is start with the preparation of a calendar of science observations using the COPT to produce an initial TOL for the OOPT. The Operator then produces the initial OPL, and verifies the possibility to execute the Plan with the Script Simulator. After harmonization with SOC, the PTR is produced and harmonized. Finally, the ITL is produced and harmonized. This concludes with the production of the PORs with the computation of the script parameters for each observation. At every intermediate step, the OOPT can produce reports, plots, etc.}
    \label{fig:OPT-flowchart} 
  \end{figure}

  The preparation of the planning and uplink file deliveries by the SWI Team follow slightly different process for ongoing Cruise Phase operations and the future Science Phase after the JOI. This distinction is driven by external procedural approaches of MOC and SOC that are required during the different phases. Nevertheless, for SWI there is a strong overlap, also in the software used to produce the necessary commanding files and meta-data. In general, SWI planning relies on the selection of observations based on science priorities, and on the subsequent computation of the instrument parameters for each of the selected observation and corresponding observation mode. This is achieved with a suite of observation planning tools, which are described in the following sections. They constitute the SWI Commanding pipeline. The full flowchart summarizing the various steps from the initial inputs to the submission of PORs is shown in Fig.~\ref{fig:OPT-flowchart}.

    \subsubsection{The Co-I Observation Planning Tool (COPT)}
    The main goal of the COPT is to provide the Science Team with the means to produce a high level schedule of the SWI observations (inflight calibration and science observations). This primarily concerns the mid-term plan (MTP). As such, the LEGA can be considered as the first inflight MTP with the whole Juice payload. The COPT is a web-based tool accessible to the instrument co-investigators (Co-Is), to allow them to prepare and propose observation schedules relevant for the MTP stage. The COPT consists of:
    \begin{MYITEMIZE}
      \item a database of observations with all the properties that define them: target, expected duration, observation mode, pointing properties, selected spectral lines, spectrometer parameters, etc. This database shall contain several hundreds to thousands observations for the complete Science Phase of the mission.
      \item a database of geometrical and event data computed from the mission SPICE kernels: target sizes, position and orientation with respect to the Juice and SWI reference frames, radial and tangential velocities relative to Juice and to the SWI boresight, etc. Events, like eclipses, transits, and occultations, are also identified at this stage.
      \item a calendar in which the observations are scheduled.
      \item a set of scheduling rules to preselect the observations that are feasible at any given point in time of the mission, and compatible with the observation requirements (e.g., spatial resolution, phase angle, particular event). These scheduling rules make use of the geometrical and event calculations performed using the SPICE kernels. 
      \item algorithms to compute the relevant instrument parameters and resources (execution time, power, data rate, and data volume) for each scheduled observation. A non-exhaustive list of parameters that are calculated, and subsequently appended to the invocation of an observation mode in a TC, comprises: the tuning indexes for one, the other or both bands; the number of spectral channels (for the ACS) or the start and end channels (for the CTS); a spectral binning factor (for the CTS); activation switch of the CTS comb (for the calibration of the frequency scale); choice of mechanism end-switches for pointing initialization; number of AT and CT steps to the sky position, to the target and to the external hot load (when applicable); mapping or scanning AT and CT steps; frequency throw (for frequency-switching modes); ON and OFF integration times for the CTS or the ACS and/or the CCH. Finally, loop counters and timing delays between subsequent commands are computed to match at best the expected observation duration, which is derived from observation simulations (so as to reach a given signal-to-noise ratio). It also produces the list of required metadata, that includes: version of SPICE kernels, observation start time stamp, total execution time, total power, total energy, number of measurements (CTS, ACS and CCH), data volume, data rate, distance to the Sun, target identifier, target geometrical properties, etc. An external observation geometry visualization tool is then used to check if an unwanted source can contaminate the OFF position of an observation (see examples in section \ref{sec:Earth-post-LEGA-obs}).
    \end{MYITEMIZE}
    The output produced by the COPT is list of time stamps, observation mode invocations with appropriate list of parameters and corresponding metadata. This list can also be exported in a format of what we termed a time-ordered-list (TOL) of SWI operations, that can be later imported into the Operator Observation Planning Tool (OOPT -- see next section; the COPT can also import a TOL file produced by the OOPT). With this information the SWI science operation manager and the operators enter into the ``harmonization'' process which iterates with MOC (during the Cruise Phase) or SOC (during the Science Phase) to account for external spacecraft or other instrument constraints on, for example, power, pointing, data volume and etc. to achieve the final scheduling of SWI observations. Because the number of scheduled SWI observations, each of which being unique due to the input parameters, can reach up to several hundreds, the final harmonization is achieved with another software, the Operators Observation Planning Tool (OOPT).

    \subsubsection{The Operator Observation Planning Tool (OOPT)}
    The OOPT is a software designed to assist with the MOC/SOC harmonization process and ultimately produce the various files necessary for the preparation of the delivery of the planning files to ESA (MOC or SOC). This tool comprises the following features:
    \begin{MYITEMIZE}
        \item A helper procedure to fully populate the TOL that uses a set of predefined scheduling rules (``Decision Tree'' in Fig.~\ref{fig:OPT-flowchart}), the segmentation proposed by SOC, and the SPICE kernels. It produces a generic TOL for the purpose of preparing the Long-Term Plan (LTP). This TOL can be imported into the COPT to provide scheduling guidelines to the Science Co-Is for further refinement. 
        \item A TOL editor that the operator can use during the various harmonization stages to adapt the timeline of operations.
        \item A series of functionalities that enable plotting and reporting on the various operations of the TOL.
        \item A script simulator, which purpose is to provide the following information on what a given observation mode/script will do with the provided parameters:
        \begin{MYITEMIZE}
          \item check the entire execution with timings and check that the script will execute properly
          \item accumulate AT, CT and FLM motor movements, information that is also stored for lifetime monitoring
          \item provide instantaneous AT and CT motor steps, which are converted to angular offsets and define the planned SWI pointing. 
          \item provide instantaneous as well as accumulated execution time of a script for double checking against COPT results
          \item estimate average data rate (DR) and the total expected data volume (DV).
        \end{MYITEMIZE}

    \end{MYITEMIZE}
    Furthermore, it is during the OOPT process that a unique ObsID is assigned to each of the scheduled observations. 

    \subsubsection{Conclusion on the SWI OPTs}  
    Both the OOPT and COPT, along with their generated products, are continuously being validated with the ongoing Cruise Phase operations. This is particularly true in dedicated SWI pointing observations conducted during the Near-Earth Commissioning Phase, the initial Payload Checkout Windows, and the LEGA campaigns. These activities have been instrumental in refining and achieving fully functional, integrated OPTs across all observation modes.

\section{Observation strategy for the LEGA \label{sec:obs_strategy}}

  \subsection{Context}
\begin{figure}[!t]
    \begin{center}
      \includegraphics[width=12cm]{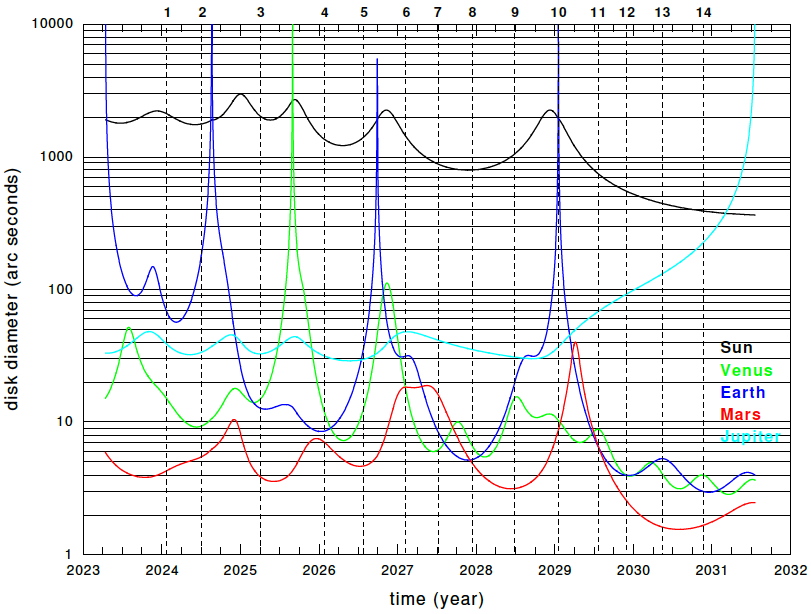}
    \end{center}
    \caption{Angular sizes of the Sun (black line), Venus (green line), the Earth (blue line), Mars (red line), and Jupiter (cyan line), as seen from Juice during the Cruise Phase. The different PCWs are labeled on the upper x-axis at their corresponding epoch.}
    \label{fig:Cruise_Phase} 
  \end{figure}

  The Cruise Phase of Juice began right after its successful launch on April 14, 2023. It will last for about 8 years, until its arrival at Jupiter in July 2031. The evolution of angular sizes of the Sun, Venus, the Earth, Mars, and Jupiter, as seen from Juice during the Cruise Phase, are shown in Fig.~\ref{fig:Cruise_Phase}. The first three months were used to commission the spacecraft and its instruments, during the NECP. The subsequent long interplanetary phase leading to Jupiter Orbital Insertion (JOI) comprises several regular Payload Checkout Windows, meant for periodic functional checks of instruments, performance verification and calibration. It also involves four planetary flybys to send Juice to Jupiter \citep{Boutonnet2024}. These gravity assists enable controlled deviations in the spacecraft’s trajectory, ultimately placing it into a heliocentric orbit tangential to Jupiter’s orbit by July 2031, ensuring a successful JOI.
  
  From the planning and operational point of view the Earth gravity assists manoeuvres provide great analogues for future Galilean moon flybys, testing SWI's relevant operational timelines. The first of these encounters, the LEGA, occurred in August 2024 and was actually a double gravity assist, with a flyby of the Moon on August 19, 2024, with a closest approach altitude of about 700\,km, followed 24 hours later by a flyby of the Earth, with a closest approach altitude of about 6800\,km. Further payload operations were possible in the three days that followed the closes approach to the Earth. In total, the LEGA segment resulted in the execution of a total of 166 observations by SWI, for 225 observations planned. Some observations were not executed following an instrument safe mode event that is explained in section \ref{sec:Earth-post-LEGA-obs}. 
  
  The general strategy for the combined flybys was three-fold: (i) to validate in representative close flyby operating conditions the use of several key observation modes, (ii) to perform spectral, and spatial calibration observations using SWI's own pointing utilizing known strong lines and/or continuum, and (iii) search for new lines in the Moon exosphere and in the Earth atmosphere. It should be noted that it was the first time the atmosphere of the Earth was observed from space at frequencies around 1200\,GHz. A general overview of the Moon and Earth observations is provided in the next sections.

  \subsection{Moon flyby}
\begin{figure}[!t]
    \begin{center}
      \includegraphics[width=18cm]{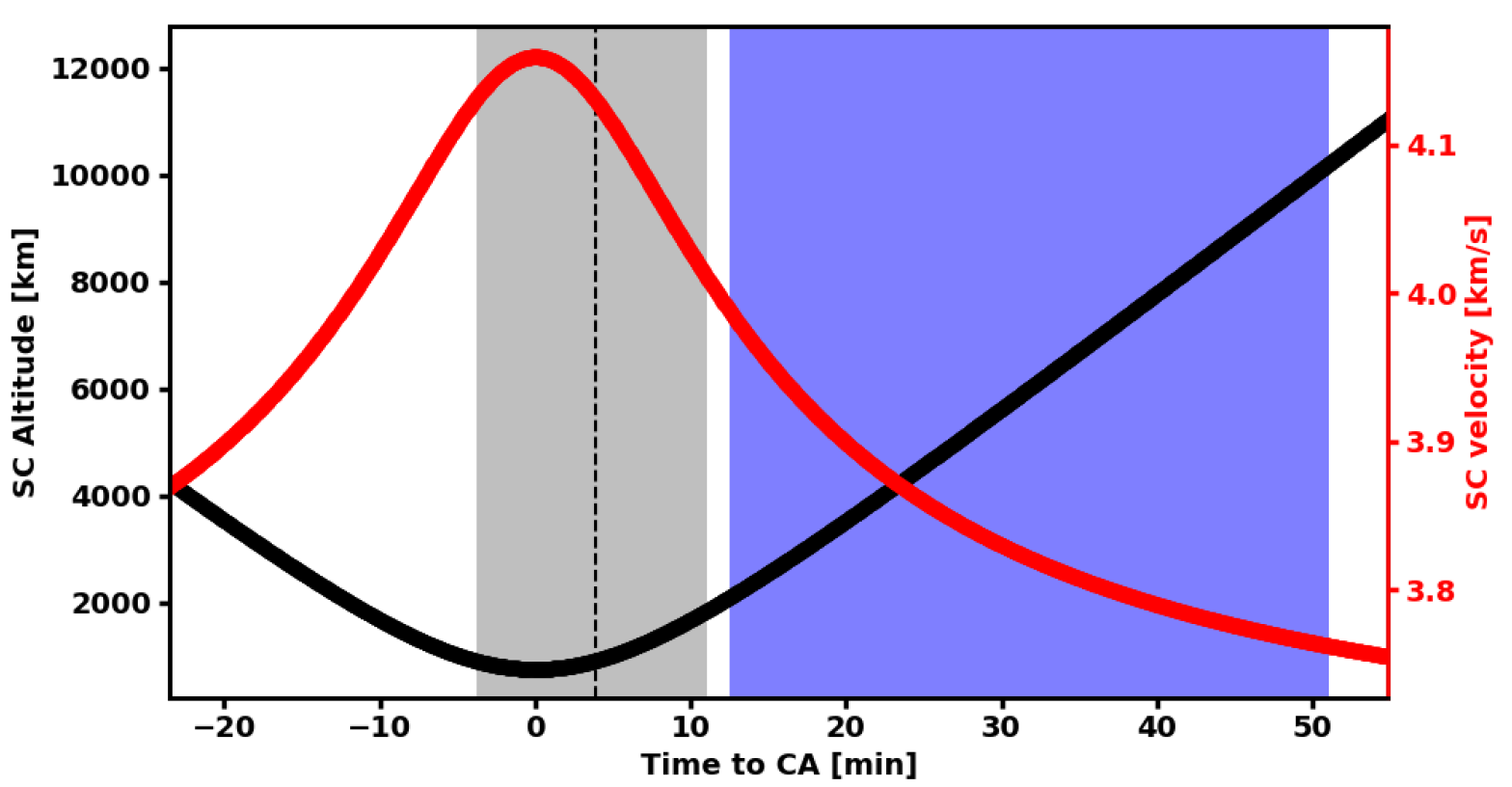}
    \end{center}
    \caption{Juice altitude in km (black line) and relative velocity in km s$^{-1}$ (red line) around the closest approach during the Moon flyby, on August 19, 2024. The gray and blue (respectively) areas represent the time windows during which the ObsIDs 227 and 228 (respectively) were recorded, i.e., when SWI had the Moon in its FOV. The black dashed line indicates the terminator crossing, and the gap between the two filled areas shows the time required to switch from the first to the second observation and move the SWI antenna back to the Moon with the AT. }
    \label{fig:Moon_flyby} 
  \end{figure}

\begin{figure}[!t]
    \begin{center}
      \includegraphics[width=15cm]{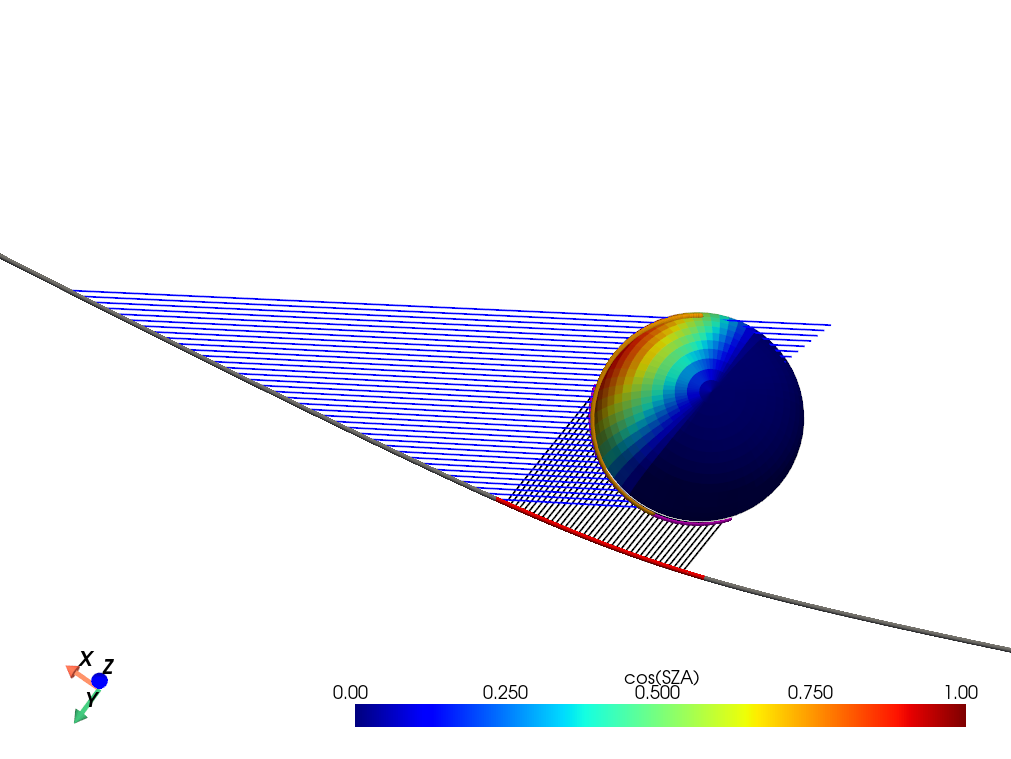}
    \end{center}
    \caption{Modeled view of the Moon flyby. Colored facets indicate the cosine of the solar zenith angle (SZA). Dark red corresponds to the sub-solar point and dark blue to the anti-solar point. The observer is positioned above the north pole and the spacecraft trajectory, which is shown as a gray line with red dots for times when SWI FOV intersected the Moon surface during the first observation (ObsID 227) and direction of SWI FOV is shown as black lines. The blue lines represent the SWI boresight direction for the second observation (ObsID 228). The spacecraft flies from right to left (from night-side to day-side) and the velocity vector is following the gray line in that direction. Therefore, from this view angle one can see that the angle between the spacecraft velocity vector and the SWI FOV is close to 90$^\circ$$^\circ$  for the first observation (actually in the range from 110$^\circ$ to 115$^\circ$) and even larger (approximately 160$^\circ$$^\circ$ ) for the second one.}
    \label{fig:Moon_flyby_track} 
  \end{figure}

  The key characteristics of the Moon flyby in the context of SWI pointing are shown in Figs.~\ref{fig:Moon_flyby} and \ref{fig:Moon_flyby_track}. The spacecraft attitude commanding  around $\pm$50\,min the closest approach were restricted to keep optimal conditions for the geophysics (in-situ) instruments. The +Z axis of Juice was fixed essentially in a nadir view during this period. Because the lunar distance changed very rapidly near closest approach, it was not feasible to perform extensive mapping of the Moon’s emission across different latitudes and longitudes or to evaluate the SWI limb-staring, scanning, or mapping modes. Therefore, we opted for an unconventional use of the SWI\_2D\_MAP\_OTF\_V1 around closest approach, by manually setting the instrument antenna in its nadir direction (aligned with spacecraft +Z axis), commanding only a single map row, and setting mechanism offsets to zero in-between subsequent raw points so as to keep the antenna pointing unchanged. Furthermore, implementing the shortest possible integration time for the CTS, i.e., 1.5\,s, and defining a large ``map'' size has effectively enabled us to record a full track of measurements while the Moon drifted below SWI and the spacecraft during the flyby. The use of the OTF mapping mode for this observation (ObsID 227) also bore the advantage of avoiding antenna movements for the cold sky position in-between two subsequent integrations. 
      
  Once the Moon eventually moved out of the field-of-view (FOV) of the instrument, a second execution of this observation was scheduled. ObsID 228 required repointing the SWI antenna by 54$^\circ$ back (against the direction of the spacecraft velocity vector) using the AT mechanism. By doing so, SWI kept recording Moon spectra for approximately 45 additional minutes. For both ObsIDs (227 and 228), we tuned the receivers to the frequencies of the H$_2$O lines at 557\,GHz and 1113\,GHz to test those tunings, because they are crucial for the Jupiter and Galilean moon observations during the Science Phase.

  The Moon flyby observations are summarized in Table~\ref{tab:LEGA-Moon} of Appendix \ref{sec:appendix_LEGA_Moon}. 

  \subsection{Earth flyby}
\begin{figure}[!t]
    \begin{center}
      \includegraphics[width=18cm]{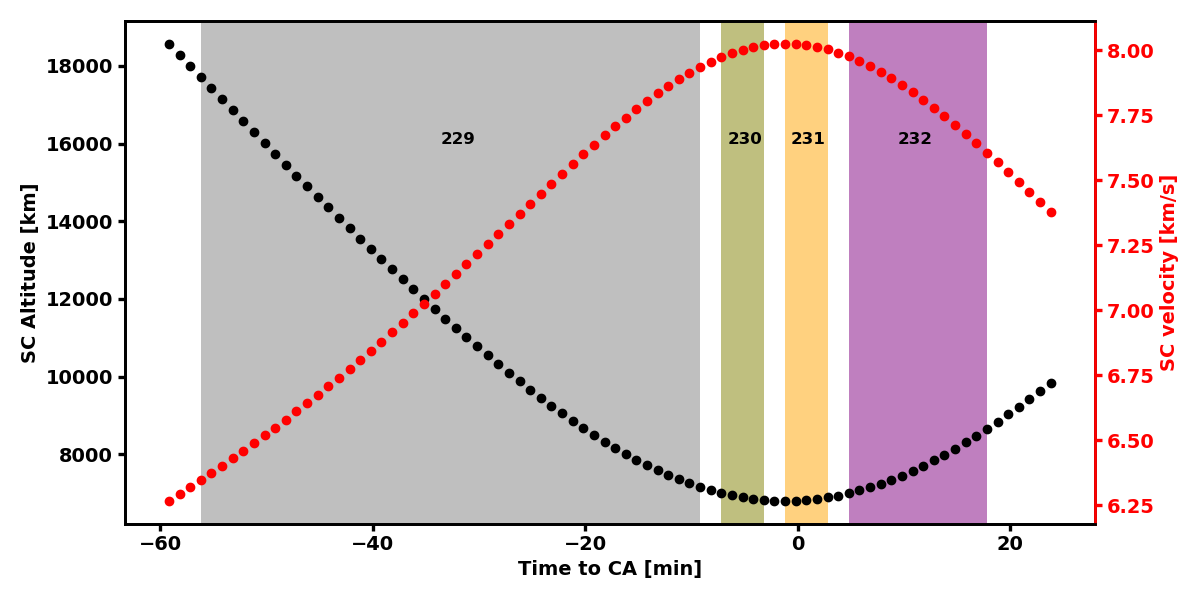}
    \end{center}
    \caption{Juice altitude in km (black dots) and relative velocity in km s$^{-1}$ (red dots) around the closest approach during the Earth flyby, on August 20, 2024. The colored areas represent the time windows during which four SWI observations were recorded. The corresponding ObsIDs are labeled. }
    \label{fig:Earth_flyby} 
  \end{figure}

  Similarly to the Moon flyby, the Juice remote sensing instrument platform (the $+$Z platform) maintained a near-nadir pointing orientation for several minutes around its closest approach to Earth. During this time SWI performed four observations (ObsIDs 229-232, see Table~\ref{tab:LEGA-Earth} in Appendix \ref{sec:appendix_LEGA_Earth}), two on either side of the closest approach. The first and fourth were similar to those executed during the Moon flyby, i.e., two strips of SWI\_2D\_MAP\_OTF\_V1 mode manual setup. Key characteristics and the timing of the four SWI observations are shown in Fig.~\ref{fig:Earth_flyby}. 
  
  The first observation used nadir pointing of the antenna, while the fourth observation (post-closest approach) required to repoint the antenna towards the Earth. In both cases, the receivers were tuned to the frequencies of the H$_2$O lines at 557\,GHz and 1113\,GHz to further test those tunings with a target atmosphere that contains H$_2$O. It should be noted that many other lines were also potentially detectable in the spectral windows covered by those two tunings. 

  The second observation, just before the closest approach, benefited from the largest relative size of the Earth atmospheric limb. Because this maximized the potential for detection of weak lines, the SWI team designed a limb scan using SWI\_MOON\_LIMB\_SCAN\_PS\_V1, with the receivers tuned to an HDO line. This observation also enabled the validation of the use of this mode in a close flyby situation. The third observation, just after the closest approach, was designed with the SWI\_MOON\_LIMB\_STARE\_PS\_V1 to observe another couple of weak lines from H$_2^{17}$O and H$_2^{18}$O, and also enabled the validation of using this mode in close flyby operating conditions. For both observations, we chose to compute the relevant pointing parameters (tangent altitude for the limb stare, altitude range and steps for the limb scan) for the mid-observation time, because of the rapidly changing size of the Earth and its limb. Because of these size changes and the inertial pointing of the spacecraft, these observations obviously suffer from pointing drift that will have to be carefully taken care of at the data analysis stage.

  \subsection{Later LEGA observations}
  A few hours after the Earth closest approach, a rotation of the spacecraft of 180$^\circ$ around its +X axis enabled SWI to have the Earth and the Moon within the reach of its mechanisms until the end of the LEGA operational phase, i.e., three days after the Earth closest approach. A comprehensive set of calibration and science observations had been prepared to be run over the three days. The observation strategy is detailed hereafter and comprises generic, Moon and Earth observations.

    \subsubsection{Generic observations}
    Every operational day started with system temperature measurements (see Table~\ref{tab:LEGA-radiometric-cal} in Appendix \ref{sec:appendix_post_LEGA_generic}) to enable a proper radiometric calibration of the data of the same day. The limitation of FLM cycling over the entire mission \citep{Hartogh2026a} resulted in only using the internal hot load to calibrate the system temperature measurements and to use those for the relative calibration of other observations using an external hot load at the disk center of the target taken the same day. All these internal/external hot load measurements will serve the purpose of constituting a long-term database for radiometric calibration of SWI data \citep{Jarchow2026} (this issue).
      
    \subsubsection{Moon observations}
    The goals of the Moon observations were three-fold. We first performed limb stares at the North, South, East and West limbs of the Moon as soon as operations resumed following the Earth gravity assist. The short time delay after the LEGA ensured that the angular size of the Moon was still large enough to consolidate the validation of the instrument pointing offsets, the limb staring mode itself, and to search for a possible water vapor exosphere \citep{Livengood2015}. These observations were recorded on August 21, 2024, from 9:50 to 11:30 UTC, and were supposed to be followed 15 hours later by an AT and a CT scan across the Moon to measure the beam in both bands. These observations could not be performed because of an instrument safe mode event a few hours earlier, during the execution of ObsID 314 (see next section). After the recovery of the instrument, final Moon observations were carried out in the form of various spectral scans all in nadir (both using the CTS and the ACS, and both in position-switch and frequency-switch) on August 23, 2024, to enable the characterization of the spectral response and constrain baseline standing waves across the full bandwidth. A high-level summary of these observations can be found in Table~\ref{tab:LEGA-Moon2} of Appendix \ref{sec:appendix_post_LEGA_Moon}.
      
    \subsubsection{Earth observations \label{sec:Earth-post-LEGA-obs}}
    Numerous Earth observations were made from August 21 to 23, 2024. A high-level summary of these is provided in Table~\ref{tab:LEGA-Earth2} of Appendix \ref{sec:appendix_post_LEGA_Earth}. In the subsequent discussion, we categorize our approach according to the operational, calibration, or scientific objective. An additional obvious objective was to validate the chosen observation modes.

\begin{figure}[!t]
      \begin{center}
        \includegraphics[width=16cm]{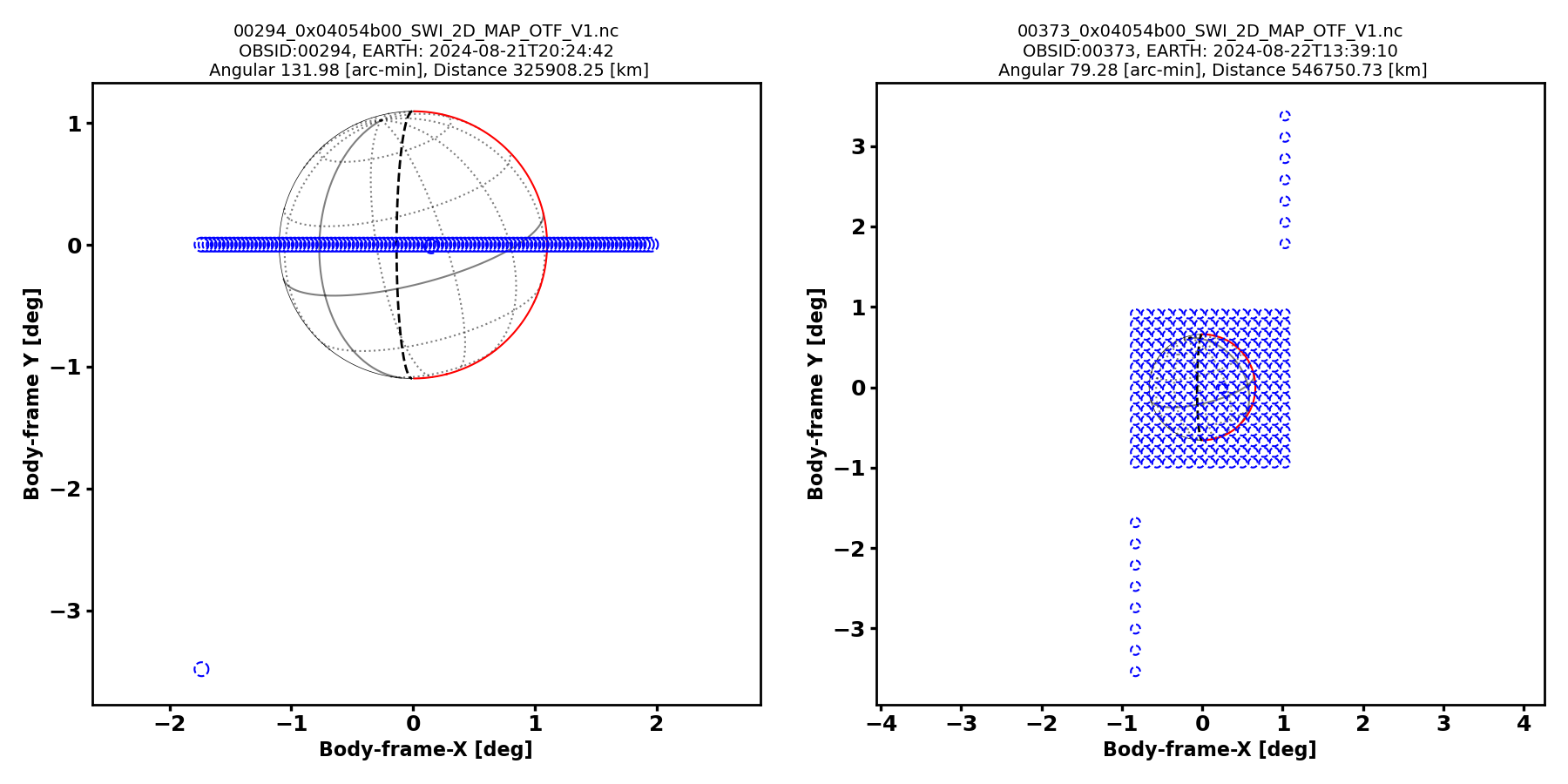}
      \end{center}
      \caption{Geometry of the Earth for (Left) the AT scan performed on August 21, 2024, and (Right) one of the 2D OTF maps performed on August 22, 2024, during the LEGA phase. The Earth is represented by the gray circle, with its equator also shown with a gray line. Pointed positions depicted by the blue dashed circles include ON source integrations, the OFF integration (the furthest to the bottom left), and the “external” hotload calibration (slightly right to the disk center). The South Pole is at the top. The first observation served to validate the SWI\_2D\_MAP\_OTF\_V1 mode used with a single row. Another observation that used a single column was also successful.}
      \label{fig:LEGA_Earth_AT_scan_2D_map} 
    \end{figure}

\begin{figure}[!t]
      \begin{center}
        \includegraphics[width=16cm]{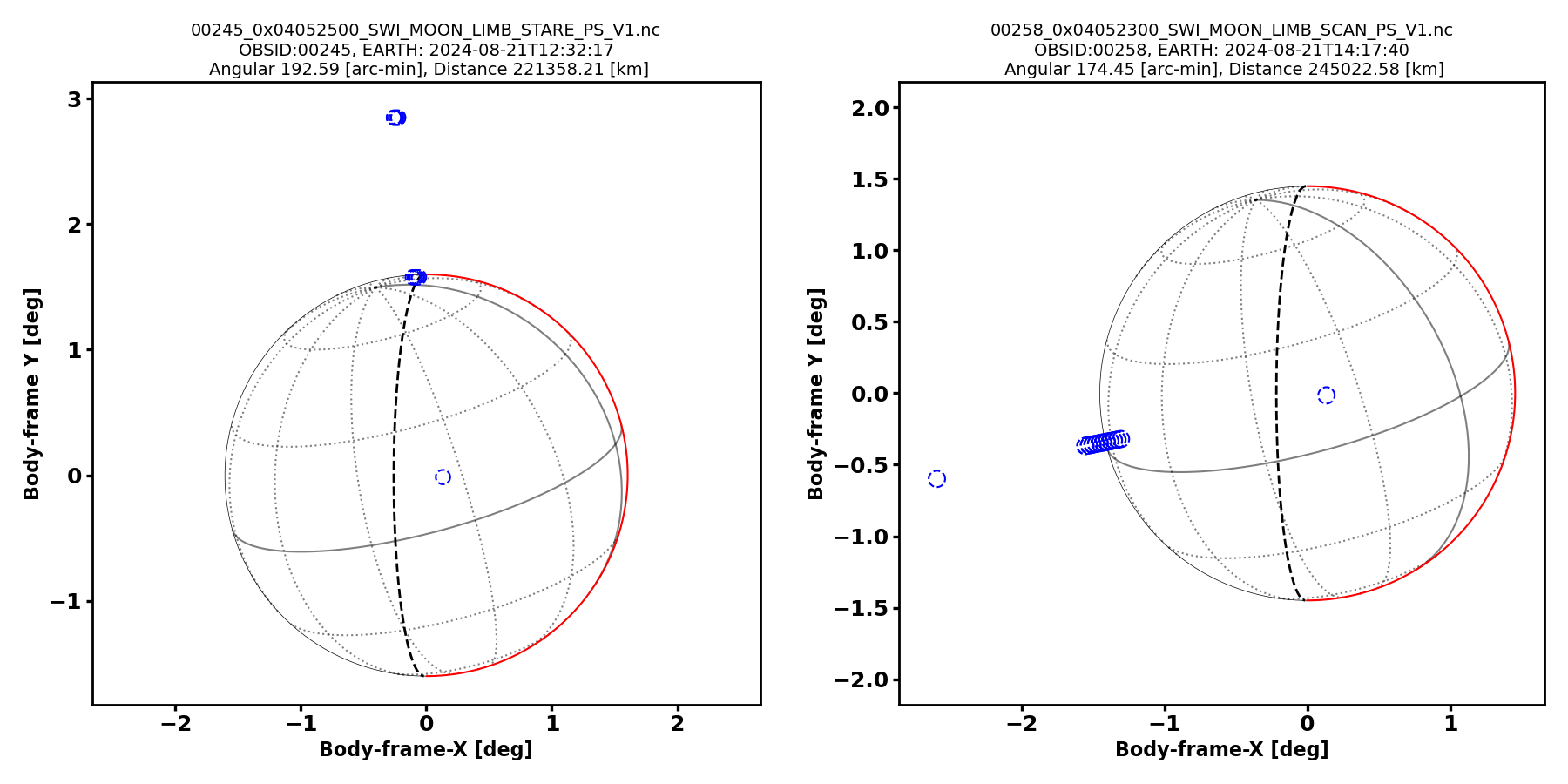}
      \end{center}
      \caption{Geometry of the Earth during (Left) the North Pole limb stare and (Right) the equatorial nightside limb scan, both performed on August 21, 2024, during the LEGA phase. Same layout as Fig.~\ref{fig:LEGA_Earth_AT_scan_2D_map}. These observations served to validate the SWI\_MOON\_LIMB\_STARE\_PS\_V1 and SWI\_MOON\_LIMB\_SCAN\_PS\_V1 modes.}
      \label{fig:LEGA_Earth_limbs} 
    \end{figure}

    Using the SWI\_2D\_MAP\_OTF\_V1 mode, regular 1D scans were performed either along the AT or CT direction plus/minus a few beams to reach cold space, as shown in Fig.~\ref{fig:LEGA_Earth_AT_scan_2D_map} (left)\footnote{The planet rendering in Figs.~\ref{fig:LEGA_Earth_AT_scan_2D_map} and \ref{fig:LEGA_Earth_limbs} are made with the help of PlanetMapper python package \citep{King2023}, available at https://github.com/ortk95/planetmapper.}. Also, several 15$\times$15 point maps were planned on August 22 (see Fig.~\ref{fig:LEGA_Earth_AT_scan_2D_map} right), with 12 different tunings to cover a variety of lines. Only the last 8 maps were actually performed, after the Mission Operation Center (MOC) was able to restart the instrument's operations timeline following the safe mode triggered by ObsID 314 (see below). The goal of these maps was to characterize the beam pattern and verify the pointing accuracy. Examples of these observations are provided as video supplements with the following DOIs: \href{https://doi.org/10.5446/72321}{10.5446/72321}, \href{https://doi.org/10.5446/72322}{10.5446/72322}, and \href{https://doi.org/10.5446/72325}{10.5446/72325}, for ObsID 242, 243, and 373, respectively. This inflight calibration of the antenna pointing and beam characterization is presented in \citet{Moreno2026} (this issue). Additional 1D scans were performed along the equator and along the central meridian. Here, we aimed at validating our commanding for proper tilting of 2D maps in order to align them with the rotation axis of a planet. In practice, the map point coordinates (and consequently the map point AT/CT steps) are rotated by the North Pole angle, which is obtained from the SPICE kernels. This planning feature will be important especially when mapping Jupiter, to align the maps with the rotation axis of the planet and therefore facilitate the planet rotation correction on the spectra of each map point.

    Multiple limb stares and limbs scans were performed either with position- or frequency-switching, with the double objective of consolidating the knowledge of the instrument pointing (with respect to the Juice nadir direction) and to confirm several tunings crucial for the Science Phase. To make sure a minimum of useful data would be acquired, Earth-specific tunings with strong O$_3$ and O$_2$ lines were also selected among the various one planned in these observations. Two examples of such observations are in Fig.~\ref{fig:LEGA_Earth_limbs} and two others are provided as video supplements with the following DOIs: \href{https://doi.org/10.5446/72323}{10.5446/72323} and \href{https://doi.org/10.5446/72324}{10.5446/72324}, for ObsID 257 and 282, respectively. 

    To further validate the variety of spectral tunings that will later be used during the Science Phase, we scheduled a number of spectral scans either with the CTS or the ACS, both, in position- and frequency-switching mode. In total, we tested 75 selected tunings in each spectral band. These spectral scans are presented in \citet{Jarchow2026} (this issue). The pointing geometry of the different  spectral scans includes nadir and various limb staring positions (see Table~\ref{tab:LEGA-Earth2}). Some of the planned spectral scans were eventually not executed because of the safe mode caused by ObsID 314. This safe mode was triggered by insufficient tuning stabilization time before the start of signal acquisition with the CTS. This has been tested and confirmed during PCW\#03 in 2025 and longer tuning stabilization times have been applied ever since.

\section{Conclusions}\label{sec:Conclusion}
This paper details the operational, and strategic plan of Juice's SWI instrument during the LEGA mission phase. We also bring into forefront the inherent complexity of this submillimetre mini-telescope, which necessitated extensive validation campaigns and calibration activities. In order to reach the SWI science goals, 31 distinct observation modes were designed, each with its own calibration approach. Currently, 27 of the 31 modes have been validated for operational use, enabling robust scientific observations in the subsequent Science Phase. In the context of SWI operations and planning, our future efforts will focus on validating and refining these modes in the subsequent Earth flybys, EGA \#2 and EGA \#3, which scheduled as part of Juice tour on its way to Jupiter.

The LEGA operations of SWI have produced a wealth of data, with 166 successful observations, which have been partly explored and analyzed \citep{Jarchow2026,Moreno2026} (this issue). It should be noted that the power and telemetry budgets of the SWI were monitored during the entire LEGA activities and were found to conform with predictions. The data acquired during the LEGA will serve to set the foundations of the future Earth Gravity Assists, namely EGA \#2 on 28 September 2026 and EGA \#3 on 17 January 2029. The instrument highest priorities for EGA \#2 and \#3 will undoubtedly include the consolidation of the knowledge of uncertainties of the two axes of the pointing mechanism, both absolute and relative. Another goal will be to consolidate the flyby operational strategy, in view of the future Europa, Ganymede and Callisto flybys. A strategy broadly similar to that applied during the LEGA, with AT/CT scans, limb stare/scans, spectral scans, and 2D OTF maps, may thus apply for these EGAs.

\begin{appendices}
    \section{Calibration modes \label{sec:appendix_calibration}}
    \begin{MYITEMIZE}
      \item \textbf{SWI\_TSYS\_CTS\_V1}: This mode consists of a spectral scan to measure the system temperature spectra of the 2 bands with the CTS 1 \& 2 by observing the hot load and cold sky. Integration time on the CTS is 2 seconds. A single execution can cover up to 15 tunings. The pointing of the spacecraft can be any and the instrument does not use its pointing mechanism.
      
      \item \textbf{SWI\_TSYS\_ACS\_CCH\_V1}: This mode is similar to SWI\_TSYS\_CTS\_V1, but using the ACS \& CCH 1 \& 2. Integration time on ACS is 1 second. A single execution can cover up to 15 tunings. An initial version this mode, named \textbf{SWI\_TSYS\_ACS\_CCH} is no longer used and will be removed from the DPU in a future PCW.

      \item \textbf{SWI\_TSYS\_ACS\_V1}: This mode is similar to SWI\_TSYS\_ACS\_CCH\_V1, but CCH 1 \& 2 are switched off. A single execution can cover up to 16 tunings.

      \item \textbf{SWI\_TSYS\_CCH\_V1}: This mode is similar to SWI\_TSYS\_ACS\_CCH\_V1, but ACS 1 \& 2 are switched off. A single execution can cover up to 16 tunings.

      \item \textbf{SWI\_ALLAN\_CTS\_FS}: Allan variance characterization of the CTS 1 \& 2 by integrating on the cold sky. Integration time is 1.5\,s and the calibration method is frequency-switch. The pointing of the spacecraft can be any and the instrument does not use its pointing mechanism. 

      \item \textbf{SWI\_ALLAN\_ACS\_FS}: This mode is similar to SWI\_ALLAN\_CTS\_FS, but using the ACS 1 \& 2. Integration time is 1\,s.

      \item \textbf{SWI\_ALLAN\_TOTAL\_CTS}: Total power Allan variance characterization of the CTS 1 \& 2 by integrating on the cold sky. Integration time is 1.5\,s. The pointing of the spacecraft can be any and the instrument does not use its pointing mechanism.
      
      \item \textbf{SWI\_ALLAN\_TOTAL\_ACS}: This mode is similar to SWI\_ALLAN\_TOTAL\_CTS, but using the ACS 1 \& 2. Integration time is 1\,s. 
      
      \item \textbf{SWI\_ALLAN\_TOTAL\_CCH}: This mode is similar to SWI\_ALLAN\_TOTAL\_CTS, but using the CCH 1 \& 2. Integration time is 1\,s. 

    \end{MYITEMIZE}
    It should be noted that all integration times are only given for typical use cases, but are parameters when invoking a mode and can thus be tuned to the desired value.

    \section{Science observation modes \label{sec:appendix_science}}
    \begin{MYITEMIZE}
      \item \textbf{SWI\_NADIR\_STARE\_PS\_V1}: Staring observations in the nadir position of the instrument (the spacecraft pointing can either be nadir or limb) to investigate atmospheric composition and temperature of Jupiter and the Galilean moons (see Fig.~\ref{fig:Nadir_stare} left). This mode is nominally meant for deep integrations and requires numerous repetitions. Nominally, two CTS spectra are recorded for 60 seconds over 10000 channels using a position-switch calibration method. An initial version this mode, named \textbf{SWI\_NADIR\_STARE\_PS} is no longer used and will be removed from the DPU in a future PCW.
      
    \begin{figure}[!t]
      \begin{center}
        \includegraphics[width=16cm]{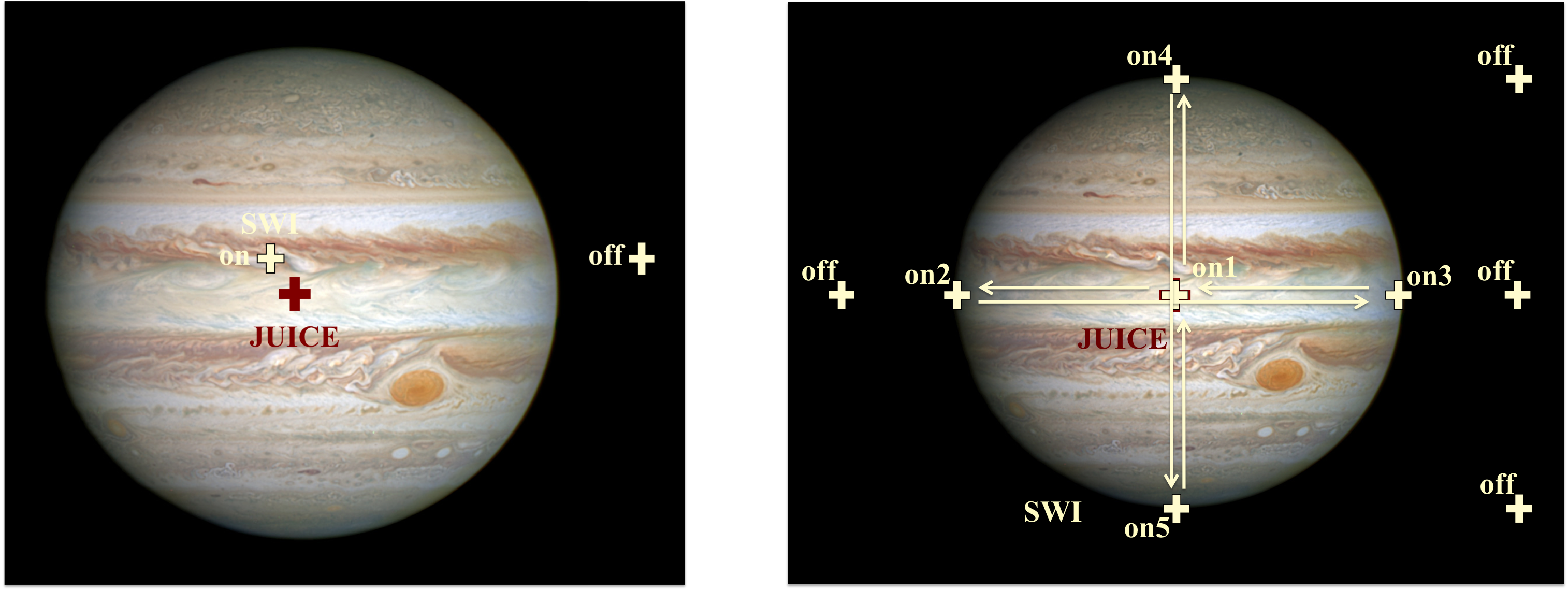}
      \end{center}
      \caption{(Left) SWI\_NADIR\_STARE\_* or SWI\_MOON\_NADIR\_STARE* (and also SWI\_SPECTRAL\_SCAN\_ACS\_* and SWI\_SPECTRAL\_SCAN\_CTS\_*) and (Right) SWI\_5POINT\_CROSS\_* mode geometries. In frequency-switch, there is no need for OFF position pointing.}
      \label{fig:Nadir_stare} 
    \end{figure}

      \item \textbf{SWI\_NADIR\_STARE\_FS\_V1}: Same as SWI\_NADIR\_STARE\_PS\_V1, except a frequency-switch calibration mode is used instead of position-switch. It enables spending $\sim$100\% of the integration time on-source. If the purity of the spectral band is good enough, there is an option to pre-compute ON$-$OFF for the CTS before downlink.
      
      \item \textbf{SWI\_5POINT\_CROSS\_PS\_V1}: Rough raster mapping using a 5-point cross to investigation of the Jovian and Galilean moon atmospheric composition, and Galilean surface properties (see Fig.~\ref{fig:Nadir_stare} right). The stepsize is such that the opposite ends of the cross are separated by the size of the target in the given direction. For Jupiter, two CTS spectra are recorded for 60 seconds over 10000 channels. For moon monitoring, two CTS spectra are recorded for 30 seconds over 210 channels. For both cases, and in parallel, two CCH measurements are recorded every 1-2 seconds. A position-switch calibration method is used.
      
      \item \textbf{SWI\_5POINT\_CROSS\_FS\_V1}: Same as SWI\_5POINT\_CROSS\_PS\_V1, except a frequency-switch calibration mode is used instead of position-switch. It enables spending $\sim$100\% of the integration time on-source. If the purity of the spectral band is good enough, there is an option to pre-compute ON$-$OFF for the CTS before downlink. Frequency-switch calibration method for CTS data.
      
      \item \textbf{SWI\_2D\_MAP\_PS\_V1}: This is a multi-purpose mode that can be used on any science target for any 2D mapping (see Fig.~\ref{fig:2D_map}), and can be customized into meridional or zonal rasters. This mode can also be used for pointing calibration purposes. The number of rows and columns and the stepsize of the raster map is adaptable to the target angular size. For Jupiter, this mode is used for the investigation of the global and regional stratospheric composition and temperature of Jupiter, and pointing calibration. For 2D maps, meridional scans and zonal scans, two CTS spectra are recorded for 60 seconds over 10000 channels. For moon monitoring, it is used to investigate the spatial distribution of Galilean moons atmospheric species ($+$ monitoring), and calibration. Two CTS spectra are recorded for 60 seconds over 210 channels. During flybys, it is used to map Galilean Moon surface properties and atmospheric composition, temperature, and winds. Two CTS spectra are recorded for 30 seconds over 210 channels. During GCO, it is used to (1) Investigate Ganymede's atmospheric composition, temperature, and winds, and surface properties by scanning from limb to limb with the along-track mechanism across the ground-track using the antenna mechanism ($\pm$72$^\circ$) and two CTS spectra are recorded for 10 seconds over 130 channels, (2) to perform tomographic investigation of Ganymede's atmospheric and surface composition, temperature, and winds by scanning along-track from $-$30km to $+$30km of the nadir axis with 9 steps, using the rocker mechanism ($\pm$4.3$^\circ$), and with 1.5 seconds integration time for two CTS spectra over 130 channels. In all cases, two CCH measurements are recorded every 0.375-1 second. During GCO, this implies that two CCH measurements are separated by 1/2 beam at 1200\,GHz. This mode uses the position-switch calibration method (the OFF position is observed after each ON of the map is observed).
      
    \begin{figure}[!t]
      \begin{center}
        \includegraphics[width=8cm]{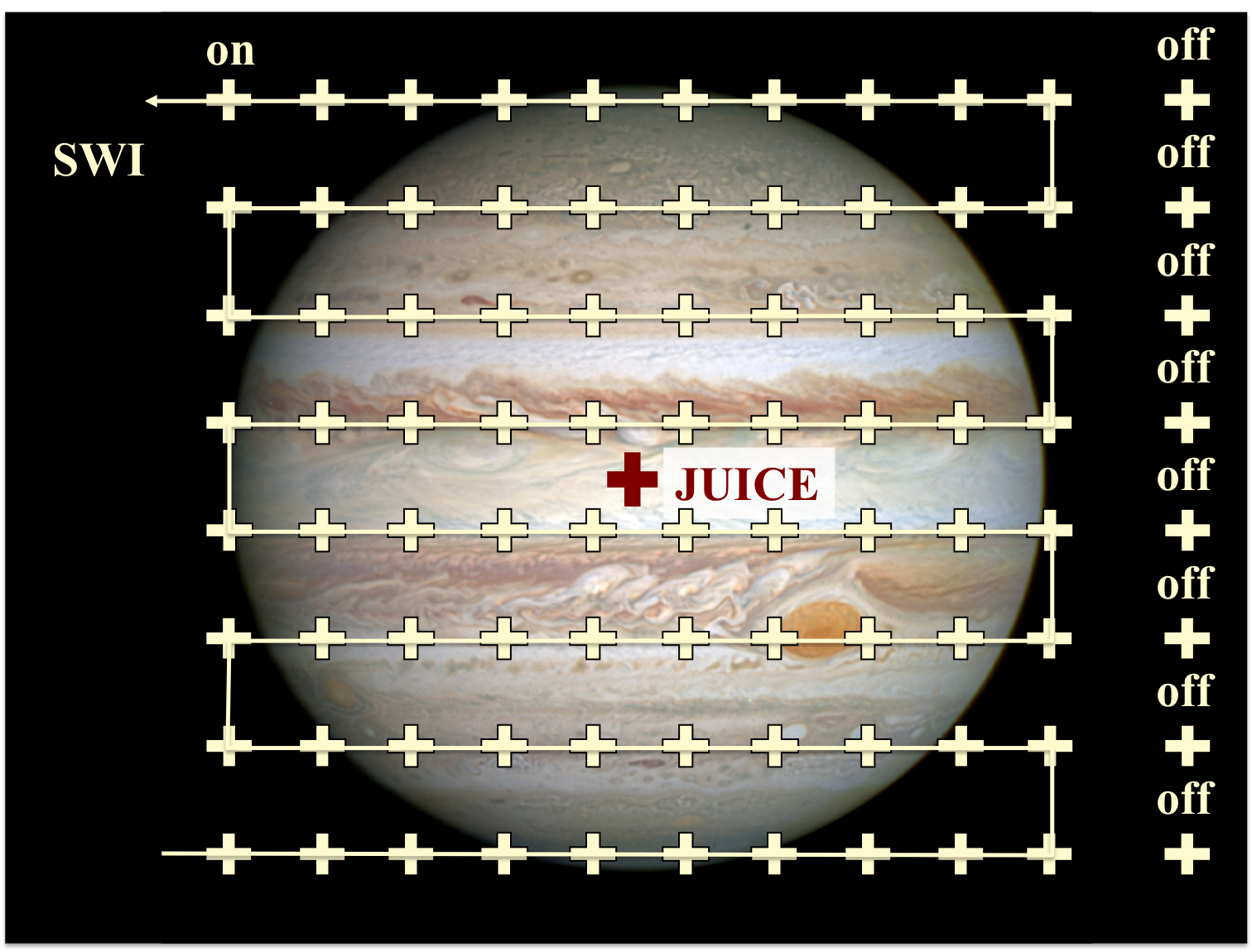}
      \end{center}
      \caption{Various SWI\_2D\_MAP\_* mode geometries. Steps on the figures are only indicative and actually depend on science goal. This generic 2D map can be used for pointing calibration, Jupiter mapping or moon monitoring. IT can be customized to a meridional raster or to a zonal raster. In frequency-switch, there is no need for OFF position pointing. In the On-the-fly (OTF) mode, the OFF position is observed only once per row.}
      \label{fig:2D_map} 
    \end{figure}

      \item \textbf{SWI\_2D\_MAP\_FS\_V1}: Same as SWI\_2D\_MAP\_PS\_V1, except a frequency-switch calibration mode is used instead of position-switch.
      
      \item \textbf{SWI\_2D\_MAP\_OTF\_V1}: Same as SWI\_2D\_MAP\_PS\_V1, but using an on-the-fly recording sequence, i.e. the OFF position per map row is only observed once. An initial version this mode, named \textbf{SWI\_2D\_MAP\_OTF} is no longer used and will be removed from the DPU in a future PCW.
      
      \item \textbf{SWI\_2D\_MAP\_OTF\_CCH\_V1}: Same as SWI\_2D\_MAP\_OTF\_V1, but this mode uses the CCH only.
      
      \item \textbf{SWI\_SPECTRAL\_SCAN\_ACS\_PS\_V1}: This mode is used for the investigation of the atmospheric composition of Jupiter and the Galilean moons using a fixed pointed position on the target, in the same way as SWI\_NADIR\_STARE\_PS\_V1 (see. Fig.~\ref{fig:Nadir_stare} left). The whole frequency range available to SWI is scanned. This mode is nominally meant for deep integrations and requires numerous repetitions. Two ACS spectra are recorded for 60 seconds over 1024 channels. Position-switch calibration method. A single execution can cover up to 16 tunings.
       
      \item \textbf{SWI\_SPECTRAL\_SCAN\_ACS\_FS\_V1}: Same as SWI\_SPECTRAL\_SCAN\_ACS\_PS\_V1, except a frequency-switch calibration mode is used instead of position-switch.  A single execution can cover up to 11 tunings.
      
      \item \textbf{SWI\_SPECTRAL\_SCAN\_CTS\_PS\_V1}: Same as SWI\_SPECTRAL\_SCAN\_ACS\_PS\_V1, except the CTS are used instead of the ACS. Two CTS spectra are recorded for 60 seconds over 10000 channels using a position-switch calibration method. A single execution can cover up to 13 tunings.
      
      \item \textbf{SWI\_SPECTRAL\_SCAN\_CTS\_FS\_V1}: Same as SWI\_SPECTRAL\_SCAN\_CTS\_PS\_V1, except a frequency-switch calibration mode is used instead of position-switch. A single execution can cover up to 9 tunings.
      
      \item \textbf{SWI\_JUP\_LIMB\_STARE\_PS\_V1}: This mode is especially designed for the investigation of Jupiter's stratospheric dynamics, composition and temperature by targeting one (or more) molecular line(s) at the planetary limb. The pointed latitude can be chosen. The retrieval of vertical profiles requires a very high signal-to-noise ratio ($\sim$100) and a very high spectral resolution (100\,kHz). A coarser spectral resolution (i.e., 500\,kHz) is sufficient for detections. This mode is nominally meant for deep integrations and implies numerous repetitions. A short $\sim$10-point across-limb scan of the continuum emission is performed with the CCH to derive a posteriori the instrument pointing and enabling accurate planet-rotation compensation (to get the zonal wind speeds), followed by two CTS spectra recorded for 60 seconds over 10000 channels (and two CCH measurements recorded every 2 seconds). The position-switch calibration method is used. This mode is presented in Fig.~\ref{fig:Limb_stare} left.
      
    \begin{figure}[!t]
      \begin{center}
        \includegraphics[width=16cm]{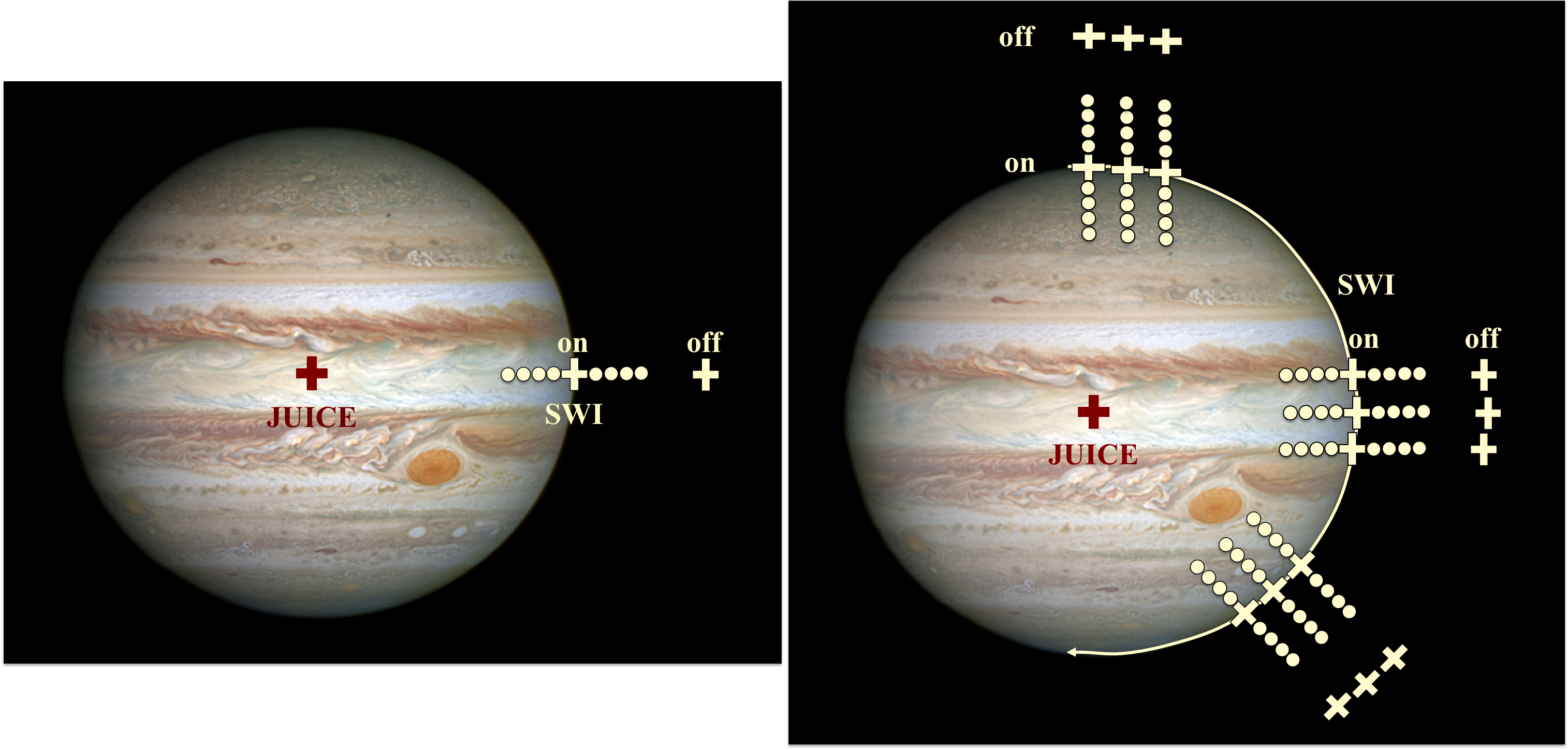}
       \end{center}
       \caption{(Left) SWI\_JUP\_LIMB\_STARE\_* mode geometry. (Right) SWI\_JUP\_LIMB\_RASTER\_* mode geometry. Each point in these modes includes a small across-limb scan to retrieved a posteriori the precise pointing of SWI. The number of points in the across-limb scan is $\sim$10, but integration times ($\sim$0.375\,s) and therefore overheads should be short. In frequency-switch, there is no need for OFF position pointing.}
      \label{fig:Limb_stare} 
    \end{figure}
    
      \item \textbf{SWI\_JUP\_LIMB\_STARE\_FS\_V1}: Same as SWI\_JUP\_LIMB\_STARE\_PS\_V1, except a frequency-switch calibration mode is used instead of position-switch.
      
      \item \textbf{SWI\_JUP\_LIMB\_RASTER\_PS\_V1}: This mode combine several limb stares from SWI\_JUP\_LIMB\_STARE\_V1 into a full series to cover a set of latitudes in a single execution. This mode is presented in Fig.~\ref{fig:Limb_stare} right).
      
      \item \textbf{SWI\_JUP\_LIMB\_RASTER\_FS\_V1}: Same as SWI\_JUP\_LIMB\_RASTER\_PS\_V1, except a frequency-switch calibration mode is used instead of position-switch. 
      
      \item \textbf{SWI\_MOON\_LIMB\_STARE\_PS\_V1}: This mode is used for the investigation of Galilean Moons atmospheric composition, temperature, and winds, by pointing to a fixed position on the limb of the target (Fig.~\ref{fig:Moon_Limb} left). The pointed latitude and altitude can be chosen. \textit{Flyby:} Two CTS spectra are recorded for 30 seconds over 210 channels. \textit{GCO:} Two CTS spectra are recorded for 30 seconds over 130 channels and a different altitude (5, 10, 20, 40, and 50 km) is scanned every orbit. this mode uses the position-switch calibration method.
      
    \begin{figure}[!t]
      \begin{center}
        \includegraphics[width=16cm]{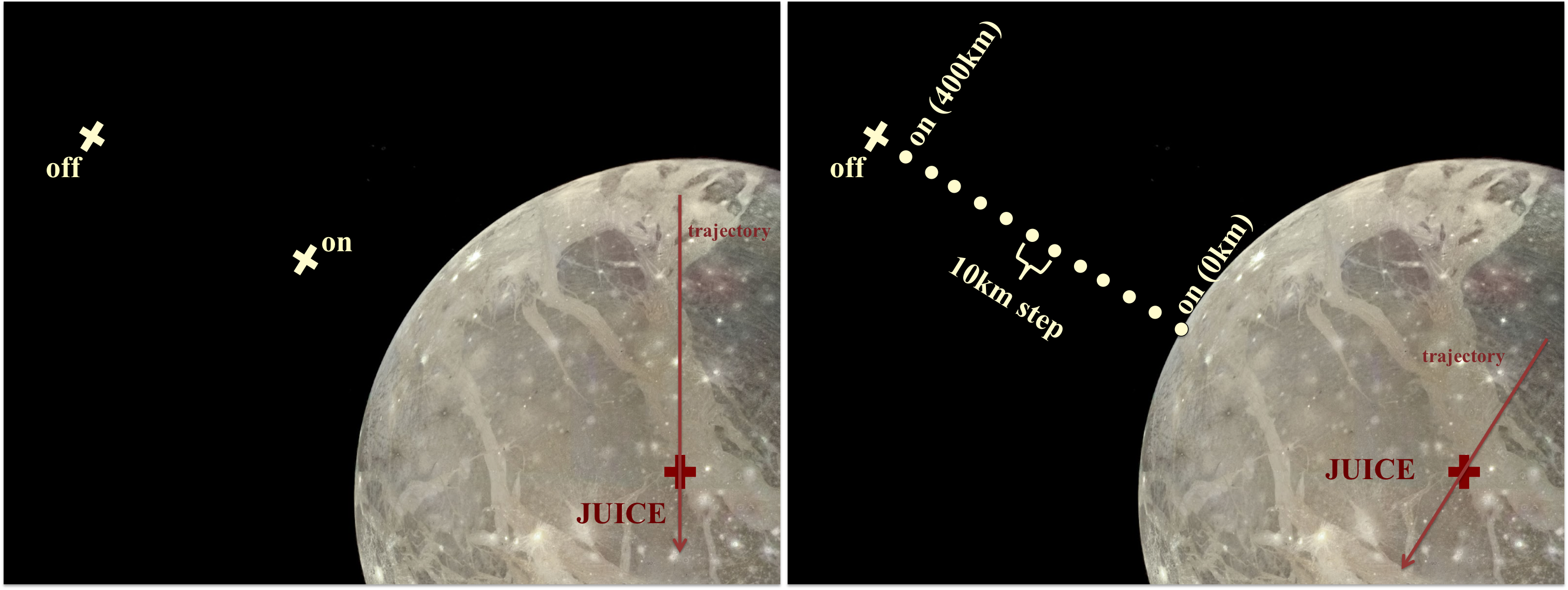}
      \end{center}
      \caption{(Left) SWI\_MOON\_LIMB\_STARE\_* mode geometry. (Right) SWI\_MOON\_LIMB\_SCAN\_* mode geometry. In frequency-switch, there is no need for OFF position pointing.}
      \label{fig:Moon_Limb} 
    \end{figure}

      \item \textbf{SWI\_MOON\_LIMB\_STARE\_FS\_V1}: Same as SWI\_MOON\_LIMB\_STARE\_PS\_V1, except a frequency-switch calibration mode is used instead of position-switch.
      
      \item \textbf{SWI\_MOON\_LIMB\_SCAN\_PS\_V1}: This mode serves to investigate the Galilean Moons atmospheric composition, temperature, and winds. \textit{Flyby:} The atmospheric limb is rapidly scanned at a chosen latitude to achieve 5\,km vertical resolution (Fig.~\ref{fig:Moon_Limb} right). Two CTS spectra are recorded for 1.5 seconds over 210 channels (16 bits coding). \textit{GCO:} The atmospheric limb is scanned up and down rapidly with 10 km altitude steps and with 1.5 seconds integration time for two CTS spectra over 130 channels. This mode uses a position-switch calibration method.
      
      \item \textbf{SWI\_MOON\_LIMB\_SCAN\_FS\_V1}: Same as SWI\_MOON\_LIMB\_SCAN\_PS\_V1, except a frequency-switch calibration mode is used instead of position-switch.
      
      \item \textbf{SWI\_MOON\_NADIR\_STARE\_PS\_V1}: This mode can be invoked for the investigation of Galilean Moons atmospheric composition, temperature, and winds, and surface properties. This mode can also be used to characterize surface polarization by pointing 45$^\circ$ off-nadir, after rotating the S/C by 90$^\circ$ around its nadir axis. It can also serve for solar occultation experiments to observe a weak molecular line in the atmosphere of Jupiter, a Galilean Moon, or the Europa torus. It uses the same concept as SWI\_NADIR\_STARE\_PS\_V1, except the CCH are also used. During flyby, two CTS spectra are recorded for 30 seconds over 210 channels. During GCO, two CTS spectra are recorded for 10 seconds over 130 channels. In both cases, two CCH measurements are recorded every 0.5-1 second, so that they are separated by maximum 1/2 beam at 1200\,GHz. During solar occultations, two CTS spectra are recorded for 60 seconds over 10000 channels, and two CCH measurements are recorded every 2 seconds. This mode uses the position-switch calibration method.
      
      \item \textbf{SWI\_MOON\_NADIR\_STARE\_FS\_V1}: Same as SWI\_MOON\_NADIR\_STARE\_PS\_V1, except a frequency-switch calibration mode is used instead of position-switch. 

    \end{MYITEMIZE}
    Similarly to the calibration modes (see Appendix \ref{sec:appendix_calibration}), all integration times are only given for typical use cases, but are parameters when invoking a mode and can thus be tuned to the desired value.

\section{Moon flyby \label{sec:appendix_LEGA_Moon}}
The two Moon observations that SWI recorded during the Moon-Gravity-Assist are detailed in Table~\ref{tab:LEGA-Moon}.

\setcounter{table}{0}
\renewcommand{\thetable}{\Alph{section}.\arabic{table}}
  \begin{table}[h]
    \caption{Observations during the Moon flyby. \label{tab:LEGA-Moon}}
    \small
    \centering
    \begin{tabular}{lrrrrr}
    \toprule
    Date & ObsID & Obs. mode & LO setup & Comment & Status \\
    \midrule
    2024-08-19T20:42:14 & 227 & SWI\_2D\_MAP\_OTF\_V1 & H$_2$O & Sub-spacecraft track & completed \\
    2024-08-19T21:26:27 & 228 & SWI\_2D\_MAP\_OTF\_V1 & H$_2$O & Backward look track &  completed \\
    \bottomrule
    \end{tabular}
  \end{table}

\section{Earth flyby \label{sec:appendix_LEGA_Earth}}
The four Earth observations performed by SWI during the Earth-Gravity-Assist are presented in Table~\ref{tab:LEGA-Earth}.

\setcounter{table}{0}
\renewcommand{\thetable}{\Alph{section}.\arabic{table}}
  \begin{table}[h]
    \caption{Observations during the Earth flyby. \label{tab:LEGA-Earth}}
    \small
    \centering
    \begin{tabular}{lrrrrr}
    \toprule
    Date & ObsID & Obs. mode & LO setup & Comment & Status \\
    \midrule
    2024-08-20T21:00:11 & 229 & SWI\_2D\_MAP\_OTF\_V1 & H$_2$O & Sub-spacecraft track & completed \\
    2024-08-20T21:49:11 & 230 & SWI\_MOON\_LIMB\_SCAN\_PS\_V1 & H$_2$O, HDO &  &  completed \\
    2024-08-20T21:55:11 & 231 & SWI\_MOON\_LIMB\_STARE\_PS\_V1 & H$_2$O, H$_2$$^{17}$O, H$_2$$^{18}$O &  & completed \\
    2024-08-20T22:01:11 & 232 & SWI\_2D\_MAP\_OTF\_V1 & H$_2$O & Backward look track &  completed \\
    \bottomrule
    \end{tabular}
  \end{table}

\section{Later LEGA observations \label{sec:appendix_post_LEGA}}
  \subsection{Generic observations \label{sec:appendix_post_LEGA_generic}}
  The radiometric calibration observations recorded after the Earth-Gravity-Assist are detailed in Table~\ref{tab:LEGA-radiometric-cal}.
  
\setcounter{table}{0}
\renewcommand{\thetable}{\Alph{section}.\arabic{table}}
  \begin{table}[h]
    \caption{Radiometric calibration observations for system temperature derivation, taken after the Earth gravity assist. Subsequent to the safe mode triggered by ObsID 314 (see Table~\ref{tab:LEGA-Moon2}), ObsIDs 322 was not executed. \label{tab:LEGA-radiometric-cal}}
    \centering
    \begin{tabular}{lrrrrr}
    \toprule
      Date                & ObsID & Obs. mode               & LO setup   & Status \\
    \midrule
      2024-08-21T11:30:12 & 237   & SWI\_TSYS\_CTS\_V1      & 15 tunings & completed \\
      2024-08-21T11:36:12 & 238   & SWI\_TSYS\_CTS\_V1      & 15 tunings & completed \\
      2024-08-21T11:42:12 & 239   & SWI\_TSYS\_CTS\_V1      & 15 tunings & completed \\
      2024-08-21T11:48:12 & 240   & SWI\_TSYS\_CTS\_V1      & 15 tunings & completed \\
      2024-08-21T11:54:12 & 241   & SWI\_TSYS\_CTS\_V1      & 15 tunings & completed \\
      2024-08-22T00:18    & 322   & SWI\_TSYS\_CTS\_V1      & 15 tunings & not executed \\
      2024-08-23T06:30:12 & 381   & SWI\_TSYS\_CTS\_V1      & 15 tunings & completed \\
      2024-08-23T06:36:12 & 382   & SWI\_TSYS\_CTS\_V1      & 15 tunings & completed \\
      2024-08-23T06:42:12 & 383   & SWI\_TSYS\_CTS\_V1      & 15 tunings & completed \\
      2024-08-23T06:48:12 & 384   & SWI\_TSYS\_CTS\_V1      & 15 tunings & completed \\
      2024-08-23T06:54:12 & 385   & SWI\_TSYS\_CTS\_V1      & 15 tunings & completed \\
      2024-08-23T12:30:12 & 414   & SWI\_TSYS\_ACS\_CCH\_V1 & 17 tunings & completed \\
      2024-08-23T12:36:11 & 415   & SWI\_TSYS\_ACS\_CCH\_V1 & 17 tunings & completed \\
      2024-08-23T12:42:11 & 416   & SWI\_TSYS\_ACS\_CCH\_V1 & 17 tunings & completed \\
      2024-08-23T12:48:11 & 417   & SWI\_TSYS\_ACS\_CCH\_V1 & 17 tunings & completed \\
    \bottomrule
    \end{tabular}
  \end{table}

  \subsection{Moon observations \label{sec:appendix_post_LEGA_Moon}}
  The Moon observations performed after the Earth-Gravity-Assist are summarized in Table~\ref{tab:LEGA-Moon2}.\\

\captionof{table}{Observations of the Moon after the Earth gravity assist. Subsequent to the safe mode triggered by ObsID 314, ObsIDs 323-334 were not executed. ObsIDs 412 and 413 were voluntarily skipped as a precaution because they had similar parameters as ObsID 314 and were considered a safe mode risk at the time. Eq. and $b$ stand for the Equator and the beam size in the 600\,GHz band, respectively.\label{tab:LEGA-Moon2}} \vspace{-0.6cm}
    \addtocounter{table}{-1}
    \begin{longtblr}[
      caption = {},
      ]{
      width=\textwidth,
      colspec = {lrrrrr},
      rowhead = 1,
      hline{1,2,Z} = {1pt,solid},      
      row{2-Z} = {font=\small},
      }
      Date & ObsID & Obs. mode & LO setup & Comment & Status \\
      2024-08-21T09:50:12 & 233 & SWI\_MOON\_LIMB\_STARE\_PS\_V1 & H$_2$O & Dawn, Eq. & completed \\
      2024-08-21T10:15:12 & 234 & SWI\_MOON\_LIMB\_STARE\_PS\_V1 & H$_2$O & Dusk, Eq. & completed \\
      2024-08-21T10:40:12 & 235 & SWI\_MOON\_LIMB\_STARE\_PS\_V1 & H$_2$O & South Pole & completed \\
      2024-08-21T11:05:12 & 236 & SWI\_MOON\_LIMB\_STARE\_PS\_V1 & H$_2$O & North Pole & completed \\
      2024-08-22T00:24    & 323 & SWI\_2D\_MAP\_OTF\_V1 & H$_2$O & AT scan $\pm$4$b$ & not executed \\
      2024-08-22T00:27    & 324 & SWI\_2D\_MAP\_OTF\_V1 & H$_2$O & CT scan $\pm$4$b$& not executed \\
      2024-08-22T00:30    & 325 & SWI\_MOON\_NADIR\_STARE\_PS\_V1 & H$_2$O & Moon center & not executed \\
      2024-08-22T00:34    & 326 & SWI\_MOON\_LIMB\_STARE\_PS\_V1 & H$_2$O & Dawn, Eq. & not executed \\
      2024-08-22T00:38    & 327 & SWI\_MOON\_LIMB\_STARE\_PS\_V1 & H$_2$O & Dusk, Eq. & not executed \\
      2024-08-22T00:42    & 328 & SWI\_MOON\_LIMB\_STARE\_PS\_V1 & H$_2$O & South Pole & not executed \\
      2024-08-22T00:46    & 329 & SWI\_MOON\_LIMB\_STARE\_PS\_V1 & H$_2$O & North Pole & not executed \\
      2024-08-22T00:50    & 330 & SWI\_MOON\_NADIR\_STARE\_FS\_V1 & H$_2$O & Moon center & not executed \\
      2024-08-22T01:10    & 331 & SWI\_MOON\_LIMB\_STARE\_FS\_V1 & H$_2$O & Dawn, Eq. & not executed \\
      2024-08-22T01:35    & 332 & SWI\_MOON\_LIMB\_STARE\_FS\_V1 & H$_2$O & Dusk, Eq. & not executed \\
      2024-08-22T02:00    & 333 & SWI\_MOON\_LIMB\_STARE\_FS\_V1 & H$_2$O & South Pole & not executed \\
      2024-08-22T02:25    & 334 & SWI\_MOON\_LIMB\_STARE\_FS\_V1 & H$_2$O & North Pole & not executed \\
      2024-08-23T09:42:11 & 400 & SWI\_SPECTRAL\_SCAN\_CTS\_PS\_V1 & 11 tunings & Moon center & completed \\
      2024-08-23T09:53:11 & 401 & SWI\_SPECTRAL\_SCAN\_CTS\_PS\_V1 & 11 tunings & Moon center & completed \\
      2024-08-23T10:04:11 & 402 & SWI\_SPECTRAL\_SCAN\_CTS\_PS\_V1 & 11 tunings & Moon center & completed \\
      2024-08-23T10:15:11 & 403 & SWI\_SPECTRAL\_SCAN\_CTS\_PS\_V1 & 11 tunings & Moon center & completed \\
      2024-08-23T10:26:11 & 404 & SWI\_SPECTRAL\_SCAN\_CTS\_PS\_V1 & 11 tunings & Moon center & completed \\
      2024-08-23T10:37:11 & 405 & SWI\_SPECTRAL\_SCAN\_CTS\_PS\_V1 & 11 tunings & Moon center & completed \\
      2024-08-23T10:48:11 & 406 & SWI\_SPECTRAL\_SCAN\_CTS\_FS\_V1 & 8 tunings & Moon center & completed \\
      2024-08-23T11:00:11 & 407 & SWI\_SPECTRAL\_SCAN\_CTS\_FS\_V1 & 8 tunings & Moon center & completed \\
      2024-08-23T11:12:11 & 408 & SWI\_SPECTRAL\_SCAN\_CTS\_FS\_V1 & 8 tunings & Moon center & completed \\
      2024-08-23T11:24:11 & 409 & SWI\_SPECTRAL\_SCAN\_CTS\_FS\_V1 & 8 tunings & Moon center & completed \\
      2024-08-23T11:36:11 & 410 & SWI\_SPECTRAL\_SCAN\_CTS\_FS\_V1 & 8 tunings & Moon center & completed \\
      2024-08-23T11:48:11 & 411 & SWI\_SPECTRAL\_SCAN\_CTS\_FS\_V1 & 8 tunings & Moon center & completed \\
      2024-08-23T12:00:11 & 412 & SWI\_SPECTRAL\_SCAN\_CTS\_FS\_V1 & 8 tunings & Moon center & not executed \\
      2024-08-23T12:12:11 & 413 & SWI\_SPECTRAL\_SCAN\_CTS\_FS\_V1 & 8 tunings & Moon center & not executed \\
      2024-08-23T15:48:11 & 429 & SWI\_SPECTRAL\_SCAN\_ACS\_PS\_V1 & 13 tunings & Moon center & completed \\
      2024-08-23T16:04:11 & 430 & SWI\_SPECTRAL\_SCAN\_ACS\_PS\_V1 & 13 tunings & Moon center & completed \\
      2024-08-23T16:20:11 & 431 & SWI\_SPECTRAL\_SCAN\_ACS\_PS\_V1 & 13 tunings & Moon center & completed \\
      2024-08-23T16:36:11 & 432 & SWI\_SPECTRAL\_SCAN\_ACS\_PS\_V1 & 13 tunings & Moon center & completed \\
      2024-08-23T16:52:11 & 433 & SWI\_SPECTRAL\_SCAN\_ACS\_PS\_V1 & 13 tunings & Moon center & completed \\
      2024-08-23T17:08:11 & 434 & SWI\_SPECTRAL\_SCAN\_ACS\_FS\_V1 & 10 tunings & Moon center & completed \\
      2024-08-23T17:22:11 & 435 & SWI\_SPECTRAL\_SCAN\_ACS\_FS\_V1 & 10 tunings & Moon center & completed \\
      2024-08-23T17:36:11 & 436 & SWI\_SPECTRAL\_SCAN\_ACS\_FS\_V1 & 10 tunings & Moon center & completed \\
      2024-08-23T17:50:11 & 437 & SWI\_SPECTRAL\_SCAN\_ACS\_FS\_V1 & 10 tunings & Moon center & completed \\
      2024-08-23T18:04:11 & 438 & SWI\_SPECTRAL\_SCAN\_ACS\_FS\_V1 & 10 tunings & Moon center & completed \\
      2024-08-23T18:18:11 & 439 & SWI\_SPECTRAL\_SCAN\_ACS\_FS\_V1 & 10 tunings & Moon center & completed \\
    \end{longtblr}

  \subsection{Earth observations \label{sec:appendix_post_LEGA_Earth}}
  The various Earth observations performed after the Earth-Gravity-Assist are summarized in Table~\ref{tab:LEGA-Earth2}.\\
    
\captionof{table}{Observations of the Earth after the Earth gravity assist. Subsequent to the safe mode triggered by ObsID 314 (see Table~\ref{tab:LEGA-Moon2}), Earth observations corresponding to ObsIDs 315-372 were not executed. Eq., CM and $b$ stand for the Equator, the Central Meridian and the beam size in the 600\,GHz band, respectively. \label{tab:LEGA-Earth2}} \vspace{-0.6cm}
    \addtocounter{table}{-1}
    \begin{longtblr}[
      caption = {},
      ]{
      width=\textwidth,
      colspec = {lrrrrr},
      rowhead = 1,
      hline{1,2,Z} = {1pt,solid},      
      row{2-Z} = {font=\small},
      }
      Date & ObsID & Obs. mode & LO setup & Comment & Status \\
      2024-08-21T12:00:12 & 242 & SWI\_2D\_MAP\_OTF\_V1 & H$_2$O & AT scan $\pm$10$b$ & completed \\
      2024-08-21T12:07:12 & 243 & SWI\_2D\_MAP\_OTF\_V1 & H$_2$O & CT scan $\pm$10$b$ & completed \\
      2024-08-21T12:14:12 & 244 & SWI\_NADIR\_STARE\_PS\_V1 & H$_2$O & Nadir & completed \\
      2024-08-21T12:28:12 & 245 & SWI\_MOON\_LIMB\_STARE\_PS\_V1 & H$_2$O & 73S & completed \\
      2024-08-21T12:36:12 & 246 & SWI\_MOON\_LIMB\_STARE\_PS\_V1 & H$_2$O & 50S 90W & completed \\
      2024-08-21T12:44:12 & 247 & SWI\_MOON\_LIMB\_STARE\_PS\_V1 & H$_2$O & 50S 90E & completed \\
      2024-08-21T12:52:12 & 248 & SWI\_MOON\_LIMB\_STARE\_PS\_V1 & H$_2$O & 20S 90W & completed \\
      2024-08-21T13:00:12 & 249 & SWI\_MOON\_LIMB\_STARE\_PS\_V1 & H$_2$O & 20S 90E & completed \\
      2024-08-21T13:08:12 & 250 & SWI\_MOON\_LIMB\_STARE\_PS\_V1 & H$_2$O & Eq 90W & completed \\
      2024-08-21T13:16:12 & 251 & SWI\_MOON\_LIMB\_STARE\_PS\_V1 & H$_2$O & Eq 90E & completed \\
      2024-08-21T13:24:12 & 252 & SWI\_MOON\_LIMB\_STARE\_PS\_V1 & H$_2$O & 20N 90W & completed \\
      2024-08-21T13:32:12 & 253 & SWI\_MOON\_LIMB\_STARE\_PS\_V1 & H$_2$O & 20N 90E & completed \\
      2024-08-21T13:40:12 & 254 & SWI\_MOON\_LIMB\_STARE\_PS\_V1 & H$_2$O & 50N 90W & completed \\
      2024-08-21T13:48:12 & 255 & SWI\_MOON\_LIMB\_STARE\_PS\_V1 & H$_2$O & 50N 90E & completed \\
      2024-08-21T13:56:12 & 256 & SWI\_MOON\_LIMB\_STARE\_PS\_V1 & H$_2$O & 73N & completed \\
      2024-08-21T14:04:12 & 257 & SWI\_MOON\_LIMB\_SCAN\_PS\_V1 & H$_2$O & Dayside Eq. & completed \\
      2024-08-21T14:13:12 & 258 & SWI\_MOON\_LIMB\_SCAN\_PS\_V1 & H$_2$O & Nightside Eq. & completed \\
      2024-08-21T14:22:12 & 259 & SWI\_2D\_MAP\_OTF\_V1 & H$_2$O & Eq. $\pm$4$b$ & completed \\
      2024-08-21T14:42:12 & 260 & SWI\_2D\_MAP\_OTF\_V1 & H$_2$O & CM $\pm$4$b$ & completed \\
      2024-08-21T15:02:12 & 261 & SWI\_SPECTRAL\_SCAN\_CTS\_PS\_V1 & 11 tunings & Day Eq. limb & completed \\
      2024-08-21T15:10:12 & 262 & SWI\_SPECTRAL\_SCAN\_CTS\_PS\_V1 & 11 tunings & Day Eq. limb & completed \\
      2024-08-21T15:18:12 & 263 & SWI\_SPECTRAL\_SCAN\_CTS\_PS\_V1 & 11 tunings & Day Eq. limb & completed \\
      2024-08-21T15:26:12 & 264 & SWI\_SPECTRAL\_SCAN\_CTS\_PS\_V1 & 11 tunings & Day Eq. limb & completed \\
      2024-08-21T15:34:12 & 265 & SWI\_SPECTRAL\_SCAN\_CTS\_PS\_V1 & 11 tunings & Day Eq. limb & completed \\
      2024-08-21T15:42:12 & 266 & SWI\_SPECTRAL\_SCAN\_CTS\_PS\_V1 & 11 tunings & Day Eq. limb & completed \\
      2024-08-21T15:50:12 & 267 & SWI\_SPECTRAL\_SCAN\_CTS\_PS\_V1 & 11 tunings & Day Eq. limb & completed \\
      2024-08-21T15:58:12 & 268 & SWI\_2D\_MAP\_OTF\_V1 & O$_3$, O$_2$ & AT scan $\pm$8$b$ & completed \\
      2024-08-21T16:05:12 & 269 & SWI\_2D\_MAP\_OTF\_V1 & O$_3$, O$_2$ & CT scan $\pm$8$b$ & completed \\
      2024-08-21T16:12:12 & 270 & SWI\_NADIR\_STARE\_PS\_V1 & O$_3$, O$_2$ & Nadir & completed \\
      2024-08-21T16:26:12 & 271 & SWI\_MOON\_LIMB\_STARE\_PS\_V1 & O$_3$, O$_2$ & 73S & completed \\
      2024-08-21T16:34:12 & 272 & SWI\_MOON\_LIMB\_STARE\_PS\_V1 & O$_3$, O$_2$ & 50S 90W & completed \\
      2024-08-21T16:42:12 & 273 & SWI\_MOON\_LIMB\_STARE\_PS\_V1 & O$_3$, O$_2$ & 50S 90E & completed \\
      2024-08-21T16:50:12 & 274 & SWI\_MOON\_LIMB\_STARE\_PS\_V1 & O$_3$, O$_2$ & 20S 90W & completed \\
      2024-08-21T16:58:12 & 275 & SWI\_MOON\_LIMB\_STARE\_PS\_V1 & O$_3$, O$_2$ & 20S 90E & completed \\
      2024-08-21T17:06:12 & 276 & SWI\_MOON\_LIMB\_STARE\_PS\_V1 & O$_3$, O$_2$ & Eq 90W & completed \\
      2024-08-21T17:14:12 & 277 & SWI\_MOON\_LIMB\_STARE\_PS\_V1 & O$_3$, O$_2$ & Eq 90E & completed \\
      2024-08-21T17:22:12 & 278 & SWI\_MOON\_LIMB\_STARE\_PS\_V1 & O$_3$, O$_2$ & 20N 90W & completed \\
      2024-08-21T17:30:12 & 279 & SWI\_MOON\_LIMB\_STARE\_PS\_V1 & O$_3$, O$_2$ & 20N 90E & completed \\
      2024-08-21T17:38:12 & 280 & SWI\_MOON\_LIMB\_STARE\_PS\_V1 & O$_3$, O$_2$ & 50N 90W & completed \\
      2024-08-21T17:46:12 & 281 & SWI\_MOON\_LIMB\_STARE\_PS\_V1 & O$_3$, O$_2$ & 50N 90E & completed \\
      2024-08-21T17:54:12 & 282 & SWI\_MOON\_LIMB\_STARE\_PS\_V1 & O$_3$, O$_2$ & 73N & completed \\
      2024-08-21T18:03:12 & 283 & SWI\_MOON\_LIMB\_SCAN\_PS\_V1 & O$_3$, O$_2$ & Dayside Eq. & completed \\
      2024-08-21T18:12:12 & 284 & SWI\_MOON\_LIMB\_SCAN\_PS\_V1 & O$_3$, O$_2$ & Nightside Eq. & completed \\
      2024-08-21T18:21:12 & 285 & SWI\_2D\_MAP\_OTF\_V1 & O$_3$, O$_2$ & Eq. $\pm$4$b$ & completed \\
      2024-08-21T18:39:12 & 286 & SWI\_2D\_MAP\_OTF\_V1 & O$_3$, O$_2$ & CM $\pm$4$b$ & completed \\
      2024-08-21T18:57:12 & 287 & SWI\_SPECTRAL\_SCAN\_CTS\_PS\_V1 & 11 tunings & North pole & completed \\
      2024-08-21T19:09:12 & 288 & SWI\_SPECTRAL\_SCAN\_CTS\_PS\_V1 & 11 tunings & North pole & completed \\
      2024-08-21T19:21:12 & 289 & SWI\_SPECTRAL\_SCAN\_CTS\_PS\_V1 & 11 tunings & North pole & completed \\
      2024-08-21T19:33:12 & 290 & SWI\_SPECTRAL\_SCAN\_CTS\_PS\_V1 & 11 tunings & North pole & completed \\
      2024-08-21T19:45:12 & 291 & SWI\_SPECTRAL\_SCAN\_CTS\_PS\_V1 & 11 tunings & North pole & completed \\
      2024-08-21T19:57:12 & 292 & SWI\_SPECTRAL\_SCAN\_CTS\_PS\_V1 & 11 tunings & North pole & completed \\
      2024-08-21T20:09:12 & 293 & SWI\_SPECTRAL\_SCAN\_CTS\_PS\_V1 & 11 tunings & North pole & completed \\
      2024-08-21T20:21:12 & 294 & SWI\_2D\_MAP\_OTF\_V1 & HCl, HCN & AT scan $\pm$6$b$ & completed \\
      2024-08-21T20:27:12 & 295 & SWI\_2D\_MAP\_OTF\_V1 & HCl, HCN & CT scan $\pm$6$b$ & completed \\
      2024-08-21T20:33:12 & 296 & SWI\_NADIR\_STARE\_FS\_V1 & HCl, HCN & Nadir & completed \\
      2024-08-21T20:45:12 & 297 & SWI\_MOON\_LIMB\_STARE\_FS\_V1 & HCl, HCN & 73S & completed \\
      2024-08-21T20:51:12 & 298 & SWI\_MOON\_LIMB\_STARE\_FS\_V1 & HCl, HCN & 50S 90W & completed \\
      2024-08-21T20:57:12 & 299 & SWI\_MOON\_LIMB\_STARE\_FS\_V1 & HCl, HCN & 50S 90E & completed \\
      2024-08-21T21:03:12 & 300 & SWI\_MOON\_LIMB\_STARE\_FS\_V1 & HCl, HCN & 20S 90W & completed \\
      2024-08-21T21:09:12 & 301 & SWI\_MOON\_LIMB\_STARE\_FS\_V1 & HCl, HCN & 20S 90E & completed \\
      2024-08-21T21:15:12 & 302 & SWI\_MOON\_LIMB\_STARE\_FS\_V1 & HCl, HCN & Eq 90W & completed \\
      2024-08-21T21:21:12 & 303 & SWI\_MOON\_LIMB\_STARE\_FS\_V1 & HCl, HCN & Eq 90E & completed \\
      2024-08-21T21:27:12 & 304 & SWI\_MOON\_LIMB\_STARE\_FS\_V1 & HCl, HCN & 20N 90W & completed \\
      2024-08-21T21:33:12 & 305 & SWI\_MOON\_LIMB\_STARE\_FS\_V1 & HCl, HCN & 20N 90E & completed \\
      2024-08-21T21:39:12 & 306 & SWI\_MOON\_LIMB\_STARE\_FS\_V1 & HCl, HCN & 50N 90W & completed \\
      2024-08-21T21:45:12 & 307 & SWI\_MOON\_LIMB\_STARE\_FS\_V1 & HCl, HCN & 50N 90E & completed \\
      2024-08-21T21:51:12 & 308 & SWI\_MOON\_LIMB\_STARE\_FS\_V1 & HCl, HCN & 73N & completed \\
      2024-08-21T21:57:12 & 309 & SWI\_MOON\_LIMB\_SCAN\_FS\_V1 & HCl, HCN & Dayside Eq. & completed \\
      2024-08-21T22:02:12 & 310 & SWI\_MOON\_LIMB\_SCAN\_FS\_V1 & HCl, HCN & Nightside Eq. & completed \\
      2024-08-21T22:07:12 & 311 & SWI\_2D\_MAP\_OTF\_V1 & HCl, HCN & Eq. $\pm$4$b$ & completed \\
      2024-08-21T22:23:12 & 312 & SWI\_2D\_MAP\_OTF\_V1 & HCl, HCN & CM $\pm$4$b$ & completed \\
      2024-08-21T22:39:12 & 313 & SWI\_SPECTRAL\_SCAN\_CTS\_FS\_V1 & 8 tunings & Night limb, Eq. & completed \\
      2024-08-21T22:50:12 & 314 & SWI\_SPECTRAL\_SCAN\_CTS\_FS\_V1 & 8 tunings & Night limb, Eq. & failed \\
      2024-08-21T23:01 & 315 & SWI\_SPECTRAL\_SCAN\_CTS\_FS\_V1 & 8 tunings & Night limb, Eq. & not executed \\
      2024-08-21T23:12 & 316 & SWI\_SPECTRAL\_SCAN\_CTS\_FS\_V1 & 8 tunings & Night limb, Eq. & not executed \\
      2024-08-21T23:23 & 317 & SWI\_SPECTRAL\_SCAN\_CTS\_FS\_V1 & 8 tunings & Night limb, Eq. & not executed \\
      2024-08-21T23:34 & 318 & SWI\_SPECTRAL\_SCAN\_CTS\_FS\_V1 & 8 tunings & Night limb, Eq. & not executed \\
      2024-08-21T23:45 & 319 & SWI\_SPECTRAL\_SCAN\_CTS\_FS\_V1 & 8 tunings & Night limb, Eq. & not executed \\
      2024-08-21T23:56 & 320 & SWI\_SPECTRAL\_SCAN\_CTS\_FS\_V1 & 8 tunings & Night limb, Eq. & not executed \\
      2024-08-22T00:07 & 321 & SWI\_SPECTRAL\_SCAN\_CTS\_FS\_V1 & 8 tunings & Night limb, Eq. & not executed \\
      2024-08-22T02:50 & 335 & SWI\_SPECTRAL\_SCAN\_CTS\_FS\_V1 & 8 tunings & South Pole & not executed \\
      2024-08-22T03:02 & 336 & SWI\_SPECTRAL\_SCAN\_CTS\_FS\_V1 & 8 tunings & South Pole & not executed \\
      2024-08-22T03:14 & 337 & SWI\_SPECTRAL\_SCAN\_CTS\_FS\_V1 & 8 tunings & South Pole & not executed \\
      2024-08-22T03:26 & 338 & SWI\_SPECTRAL\_SCAN\_CTS\_FS\_V1 & 8 tunings & South Pole & not executed \\
      2024-08-22T03:38 & 339 & SWI\_SPECTRAL\_SCAN\_CTS\_FS\_V1 & 8 tunings & South Pole & not executed \\
      2024-08-22T03:50 & 340 & SWI\_SPECTRAL\_SCAN\_CTS\_FS\_V1 & 8 tunings & South Pole & not executed \\
      2024-08-22T04:02 & 341 & SWI\_SPECTRAL\_SCAN\_CTS\_FS\_V1 & 8 tunings & South Pole & not executed \\
      2024-08-22T04:14 & 342 & SWI\_SPECTRAL\_SCAN\_CTS\_FS\_V1 & 8 tunings & South Pole & not executed \\
      2024-08-22T04:26 & 343 & SWI\_SPECTRAL\_SCAN\_CTS\_FS\_V1 & 8 tunings & South Pole & not executed \\
      2024-08-22T04:38 & 344 & SWI\_2D\_MAP\_OTF\_V1 & O$_3$, CO, HCN & AT scan $\pm$5$b$ & not executed \\
      2024-08-22T04:47 & 345 & SWI\_2D\_MAP\_OTF\_V1 & O$_3$, CO, HCN & CT scan $\pm$5$b$ & not executed \\
      2024-08-22T04:55 & 346 & SWI\_MOON\_LIMB\_STARE\_FS\_V1 & O$_3$, CO, HCN & 72S & not executed \\
      2024-08-22T05:04 & 347 & SWI\_MOON\_LIMB\_STARE\_FS\_V1 & O$_3$, CO, HCN & 50S 90W & not executed \\
      2024-08-22T05:13 & 348 & SWI\_MOON\_LIMB\_STARE\_FS\_V1 & O$_3$, CO, HCN & 50S 90E & not executed \\
      2024-08-22T05:22 & 349 & SWI\_MOON\_LIMB\_STARE\_FS\_V1 & O$_3$, CO, HCN & 30S 90W & not executed \\
      2024-08-22T05:31 & 350 & SWI\_MOON\_LIMB\_STARE\_FS\_V1 & O$_3$, CO, HCN & 30S 90E & not executed \\
      2024-08-22T05:40 & 351 & SWI\_MOON\_LIMB\_STARE\_FS\_V1 & O$_3$, CO, HCN & 15S 90W & not executed \\
      2024-08-22T05:49 & 352 & SWI\_MOON\_LIMB\_STARE\_FS\_V1 & O$_3$, CO, HCN & 15S 90E & not executed \\
      2024-08-22T05:58 & 353 & SWI\_MOON\_LIMB\_STARE\_FS\_V1 & O$_3$, CO, HCN & Eq 90W & not executed \\
      2024-08-22T06:07 & 354 & SWI\_MOON\_LIMB\_STARE\_FS\_V1 & O$_3$, CO, HCN & Eq 90E & not executed \\
      2024-08-22T06:16 & 355 & SWI\_MOON\_LIMB\_STARE\_FS\_V1 & O$_3$, CO, HCN & 15N 90W & not executed \\
      2024-08-22T06:25 & 356 & SWI\_MOON\_LIMB\_STARE\_FS\_V1 & O$_3$, CO, HCN & 15N 90E & not executed \\
      2024-08-22T06:34 & 357 & SWI\_MOON\_LIMB\_STARE\_FS\_V1 & O$_3$, CO, HCN & 30N 90W & not executed \\
      2024-08-22T06:43 & 358 & SWI\_MOON\_LIMB\_STARE\_FS\_V1 & O$_3$, CO, HCN & 30N 90E & not executed \\
      2024-08-22T06:52 & 359 & SWI\_MOON\_LIMB\_STARE\_FS\_V1 & O$_3$, CO, HCN & 50N 90W & not executed \\
      2024-08-22T07:01 & 360 & SWI\_MOON\_LIMB\_STARE\_FS\_V1 & O$_3$, CO, HCN & 50N 90E & not executed \\
      2024-08-22T07:10 & 361 & SWI\_MOON\_LIMB\_STARE\_FS\_V1 & O$_3$, CO, HCN & 72N & not executed \\
      2024-08-22T07:19 & 362 & SWI\_NADIR\_STARE\_FS\_V1 & O$_3$, CO, HCN & Nadir & not executed \\
      2024-08-22T07:48 & 363 & SWI\_MOON\_LIMB\_SCAN\_FS\_V1 & O$_3$, CO, HCN &  Dawn, Eq. & not executed \\
      2024-08-22T07:59 & 364 & SWI\_MOON\_LIMB\_SCAN\_FS\_V1 & O$_3$, CO, HCN &  Dusk, Eq. & not executed \\
      2024-08-22T08:10 & 365 & SWI\_MOON\_LIMB\_SCAN\_FS\_V1 & O$_3$, CO, HCN &  South Pole & not executed \\
      2024-08-22T08:21 & 366 & SWI\_MOON\_LIMB\_SCAN\_FS\_V1 & O$_3$, CO, HCN &  North Pole & not executed \\
      2024-08-22T08:32 & 367 & SWI\_2D\_MAP\_OTF\_V1 & O$_3$, CO, HCN & Eq. $\pm$4$b$ & not executed \\
      2024-08-22T08:46 & 368 & SWI\_2D\_MAP\_OTF\_V1 & O$_3$, CO, HCN & CM $\pm$4$b$ & not executed \\
      2024-08-22T09:00 & 369 & SWI\_2D\_MAP\_OTF\_V1 & H$_2$O & 15$\times$15 map & not executed \\
      2024-08-22T10:02 & 370 & SWI\_2D\_MAP\_OTF\_V1 & O$_3$, O$_2$ & 15$\times$15 map & not executed \\
      2024-08-22T11:04 & 371 & SWI\_2D\_MAP\_OTF\_V1 & HCN, HCl & 15$\times$15 map & not executed \\
      2024-08-22T12:06 & 372 & SWI\_2D\_MAP\_OTF\_V1 & O$_3$, CO, HCN & 15$\times$15 map & not executed \\
      2024-08-22T13:08 & 373 & SWI\_2D\_MAP\_OTF\_V1 & H$_2$O, HDO, CS & 15$\times$15 map & completed \\
      2024-08-22T14:10:12 & 374 & SWI\_2D\_MAP\_OTF\_V1 & ClO, H$_2$O isot. & 15$\times$15 map & completed \\
      2024-08-22T15:12:12 & 375 & SWI\_2D\_MAP\_OTF\_V1 & H$_2$O, NO, N$_2$O & 15$\times$15 map & completed \\
      2024-08-22T16:14:12 & 376 & SWI\_2D\_MAP\_OTF\_V1 & O$_3$, SO, SO$_2$ & 15$\times$15 map & completed \\
      2024-08-22T17:16:12 & 377 & SWI\_2D\_MAP\_OTF\_V1 & O$_3$, HF & 15$\times$15 map & completed \\
      2024-08-22T18:18:12 & 378 & SWI\_2D\_MAP\_OTF\_V1 & O$_3$, N$_2$O, CH$_4$ & 15$\times$15 map & completed \\
      2024-08-22T19:20:12 & 379 & SWI\_2D\_MAP\_OTF\_V1 & H$_2$O & 15$\times$15 map & completed \\
      2024-08-22T20:22:12 & 380 & SWI\_2D\_MAP\_OTF\_V1 & O$_3$, H$_2$O, NO & & \\
      2024-08-23T07:00:12 & 386 & SWI\_SPECTRAL\_SCAN\_CTS\_PS\_V1 & 11 tunings & Nadir & completed \\
      2024-08-23T07:11:12 & 387 & SWI\_SPECTRAL\_SCAN\_CTS\_PS\_V1 & 11 tunings & Nadir & completed \\
      2024-08-23T07:22:12 & 388 & SWI\_SPECTRAL\_SCAN\_CTS\_PS\_V1 & 11 tunings & Nadir & completed \\
      2024-08-23T07:33:12 & 389 & SWI\_SPECTRAL\_SCAN\_CTS\_PS\_V1 & 11 tunings & Nadir & completed \\
      2024-08-23T07:44:12 & 390 & SWI\_SPECTRAL\_SCAN\_CTS\_PS\_V1 & 11 tunings & Nadir & completed \\
      2024-08-23T07:55:12 & 391 & SWI\_SPECTRAL\_SCAN\_CTS\_PS\_V1 & 11 tunings & Nadir & completed \\
      2024-08-23T08:06:12 & 392 & SWI\_SPECTRAL\_SCAN\_CTS\_FS\_V1 & 8 tunings  & Nadir & completed \\
      2024-08-23T08:18:12 & 393 & SWI\_SPECTRAL\_SCAN\_CTS\_FS\_V1 & 8 tunings  & Nadir & completed \\
      2024-08-23T08:30:12 & 394 & SWI\_SPECTRAL\_SCAN\_CTS\_FS\_V1 & 8 tunings  & Nadir & completed \\
      2024-08-23T08:42:12 & 395 & SWI\_SPECTRAL\_SCAN\_CTS\_FS\_V1 & 8 tunings  & Nadir & completed \\
      2024-08-23T08:54:12 & 396 & SWI\_SPECTRAL\_SCAN\_CTS\_FS\_V1 & 8 tunings  & Nadir & completed \\
      2024-08-23T09:06:12 & 397 & SWI\_SPECTRAL\_SCAN\_CTS\_FS\_V1 & 8 tunings  & Nadir & completed \\
      2024-08-23T09:18:12 & 398 & SWI\_SPECTRAL\_SCAN\_CTS\_FS\_V1 & 8 tunings  & Nadir & completed \\
      2024-08-23T09:30:12 & 399 & SWI\_SPECTRAL\_SCAN\_CTS\_FS\_V1 & 8 tunings  & Nadir & completed \\
      2024-08-23T12:54:11 & 418 & SWI\_SPECTRAL\_SCAN\_ACS\_PS\_V1 & 13 tunings & Nadir & completed \\ 
      2024-08-23T13:10:11 & 419 & SWI\_SPECTRAL\_SCAN\_ACS\_PS\_V1 & 13 tunings & Nadir & completed \\ 
      2024-08-23T13:26:11 & 420 & SWI\_SPECTRAL\_SCAN\_ACS\_PS\_V1 & 13 tunings & Nadir & completed \\ 
      2024-08-23T13:42:11 & 421 & SWI\_SPECTRAL\_SCAN\_ACS\_PS\_V1 & 13 tunings & Nadir & completed \\ 
      2024-08-23T13:58:11 & 422 & SWI\_SPECTRAL\_SCAN\_ACS\_PS\_V1 & 13 tunings & Nadir & completed \\ 
      2024-08-23T14:14:11 & 423 & SWI\_SPECTRAL\_SCAN\_ACS\_FS\_V1 & 10 tunings & Nadir & completed \\ 
      2024-08-23T14:28:11 & 424 & SWI\_SPECTRAL\_SCAN\_ACS\_FS\_V1 & 10 tunings & Nadir & completed \\ 
      2024-08-23T14:42:11 & 425 & SWI\_SPECTRAL\_SCAN\_ACS\_FS\_V1 & 10 tunings & Nadir & completed \\ 
      2024-08-23T14:56:11 & 426 & SWI\_SPECTRAL\_SCAN\_ACS\_FS\_V1 & 10 tunings & Nadir & completed \\ 
      2024-08-23T15:10:11 & 427 & SWI\_SPECTRAL\_SCAN\_ACS\_FS\_V1 & 10 tunings & Nadir & completed \\ 
      2024-08-23T15:24:11 & 428 & SWI\_SPECTRAL\_SCAN\_ACS\_FS\_V1 & 10 tunings & Nadir & completed \\ 
      2024-08-23T18:32:11 & 440 & SWI\_SPECTRAL\_SCAN\_ACS\_PS\_V1 & 13 tunings & Nadir & completed \\ 
      2024-08-23T18:48:11 & 441 & SWI\_SPECTRAL\_SCAN\_ACS\_PS\_V1 & 13 tunings & Nadir & completed \\ 
      2024-08-23T19:04:11 & 442 & SWI\_SPECTRAL\_SCAN\_ACS\_PS\_V1 & 13 tunings & Nadir & completed \\ 
      2024-08-23T19:20:11 & 443 & SWI\_SPECTRAL\_SCAN\_ACS\_PS\_V1 & 13 tunings & Nadir & completed \\ 
      2024-08-23T19:36:11 & 444 & SWI\_SPECTRAL\_SCAN\_ACS\_PS\_V1 & 13 tunings & Nadir & completed \\ 
      2024-08-23T19:52:11 & 445 & SWI\_SPECTRAL\_SCAN\_ACS\_PS\_V1 & 13 tunings & Nadir & completed \\ 
      2024-08-23T20:08:11 & 446 & SWI\_SPECTRAL\_SCAN\_ACS\_PS\_V1 & 13 tunings & Nadir & completed \\ 
      2024-08-23T20:24:11 & 447 & SWI\_SPECTRAL\_SCAN\_ACS\_PS\_V1 & 13 tunings & Nadir & completed \\ 
      2024-08-23T20:40:11 & 448 & SWI\_SPECTRAL\_SCAN\_ACS\_PS\_V1 & 13 tunings & Nadir & completed \\ 
      2024-08-23T20:56:11 & 449 & SWI\_SPECTRAL\_SCAN\_ACS\_PS\_V1 & 13 tunings & Nadir & completed \\ 
      2024-08-23T21:12:11 & 450 & SWI\_SPECTRAL\_SCAN\_ACS\_PS\_V1 & 13 tunings & Nadir & completed \\ 
      2024-08-23T21:28:11 & 451 & SWI\_SPECTRAL\_SCAN\_ACS\_PS\_V1 & 13 tunings & Nadir & completed \\     
    \end{longtblr}

\end{appendices}

\section*{Author contributions}
T. Cavali\'e prepared the original manuscript. T. Cavali\'e, R. Moreno, L. Rezac, C. Jarchow, A. Schulz-Ravanbakhsh, and P. Hartogh defined the LEGA operational strategy for SWI. T. Cavali\'e and F. Herpin designed the SWI observation modes. T. Cavali\'e, A. Carrasco Gallardo, L. Rezac, S. Goodyear, and P. Mancini worked out the uplink of SWI observations. All co-authors contributed to the successful implementation of SWI and commented the manuscript.

\section*{Competing interests}
The authors declare no competing interest.

\section*{Special issue statement}
Special issue statement. This article is part of the special issue ``The first-ever lunar–Earth flyby: a unique test environment for Juice''. It is not associated with a conference.

\section*{Acknowledgements}
SWI has been designed and developed by an international consortium of institutes led by the Max Planck Institute for Solar System Research (MPS, Germany) and including the Laboratory for Studies of Radiation and Matter in Astrophysics (LERMA, France), the Space Research Centre of the Polish Academy of Sciences (CBK PAN, Poland), Chalmers University of Technology (Sweden), the Institute of Applied Physics of the University of Bern (IAP, Switzerland), the National Institute of Information and Communications Technology (NICT, Japan) and the French Space Agency CNES with additional support from the Laboratoire d'Instrumentation et de Recherche en Astrophysique of the Observatoire de Paris (LIRA, France), the Laboratoire d'Astrophysique de Bordeaux (LAB, France), the RPG Radiometer Physics GmbH (Germany), and Omnisys Instrument AV (Sweden). This development has been supported by national funding agencies and other organizations, including the Deutsches Zentrum für Luft- und Raumfahrt (DLR) and by central resources of the Max-Planck-Society. T. Cavali\'e, F. Herpin, and P. Mancini, acknowledge funding from the Centre National d'\'Etudes Spatiales (CNES). E.S. Wirstr\"om acknowledges generous support from the Swedish National Space Agency. Juice is a mission under ESA leadership with contributions from its Member States, NASA, JAXA, and the Israel Space Agency. It is the first Large-class mission in ESA's Cosmic Vision Program.

\bibliographystyle{aa}

\end{document}